\documentclass{openjournal}

\usepackage{lipsum}

\usepackage{xcolor}
\usepackage{textgreek}
\usepackage[utf8]{inputenc}
\usepackage[english]{babel}

\usepackage{hyperref}
\hypersetup{
    unicode, 
    colorlinks=true,
    linkcolor=linkcolor,
    citecolor=linkcolor,
    filecolor=linkcolor,
    urlcolor=linkcolor,
}
\usepackage{color,colortbl}
\definecolor{linkcolor}{rgb}{0.0,0.3,0.5}
\usepackage{tensind}
\tensordelimiter{?}
\DeclareGraphicsExtensions{.bmp,.png,.jpg,.pdf}
\usepackage{verbatim}
\usepackage[normalem]{ulem}
\usepackage{booktabs}
\usepackage{orcidlink}
\usepackage{soul}
\usepackage{booktabs}
\usepackage{tabularx}
\usepackage{array}

\graphicspath{ {./figs/} }

\begin{document}
\title{Detection and excision of radio-frequency interference in radio arrays}

\author{Vikram Ravi\orcidlink{0000-0002-7252-5485}$^{1,2}$}

\affiliation{
  $^{1}$Cahill Center for Astronomy and Astrophysics, MC 249-17, California Institute of Technology, Pasadena, CA 91125, USA\\
  $^{2}$Owens Valley Radio Observatory, California Institute of Technology, Big Pine, CA 93513, USA\\
  }
\thanks{Corresponding author: Vikram Ravi:\\ \href{mailto:vikram@caltech.edu}{vikram@caltech.edu}}

\begin{abstract}

The detection and excision of radio-frequency interference (RFI) is usually approached by asking how best to identify RFI within a given data product. I address a prior question: which data product is a given RFI emitter is most detectable in? I develop a model for the suppression of RFI in visibilities and tied-array beams by the antenna primary beam, near-field wavefront curvature, bandwidth smearing, and fringe tracking. The comparison between antenna-based and array-based detection reduces to a single quantity related to the mean per-baseline RFI coherence. Cross-correlation detection is preferred only above a coherence threshold of $(N-1)^{-1/2}$, where $N$ is the number of antennas. I evaluate this model for four array configurations (the Very Large Array A and D configurations, and the fixed MeerKAT and Deep Synoptic Array / DSA configurations) and five representative RFI emitters spanning distances of 5\,km to medium-Earth orbit. Compact, filled, arrays favor the cross-correlations for stationary terrestrial emitters, whereas extended sparse arrays do not. RFI emitters with significant motion relative to radio arrays, in particular aircraft and low- and medium-Earth-orbit satellites, favor the sum of antenna autocorrelations everywhere by one to seven orders of magnitude, because their coherence is destroyed by rapid fringe rotation. For RFI below the single-antenna noise, no phase-unknown cross-correlation statistic is more sensitive than the autocorrelation sum. Tied-array beams are the least sensitive data product of all: the array factor supplies no suppression of terrestrial RFI, so a beam sees an emitter as a single antenna does. I describe a real-time, GPU-based implementation of autocorrelation-based excision for the 110-antenna Deep Synoptic Array, and outline the design implications for upcoming arrays.

\end{abstract}


\maketitle

\section{Introduction}
\label{sec:intro}

Radio-frequency interference (RFI) encompasses signals detected by radio telescopes that are intentionally transmitted, typically by humans. The detection, characterization, and mitigation of RFI has been central to radio astronomy since its earliest days. Indeed, the \citet{Jansky1933} discovery of extraterrestrial, naturally occurring radio emissions occurred serendipitously during experiments to identify sources of RFI contaminating trans-Atlantic shortwave transmissions. Even in recent years, it has proven challenging to distinguish between new astronomical phenomena and uncataloged RFI sources \citep{Petroff2015,Sheikh2021}. As radio telescopes increase in sensitivity, they also operate in an increasingly crowded spectrum environment, with the proliferation of switching and digital electronic hardware, radio transmitters and communications networks, and satellites \citep{ekers2000,sihlangu2022}. The effects of RFI on radio-astronomical observations can be classified according to severity \citep[e.g.,][]{buch2025}, with levels ranging from RFI existing as unwanted signals in otherwise valid data, to RFI that causes all data to be invalidated by driving components of signal chains into non-linear operation regimes, to RFI that irreversibly damages receiver components. The latter RFI regimes require avoidance strategies, including upstream filtering and coordination strategies \citep[e.g.,][]{Hellbourg2026}. In this paper, I address the problem of identifying and excising RFI from otherwise valid digital data obtained with radio arrays. 

\begin{figure*}[t]
    \centering
    \includegraphics[width=\textwidth]{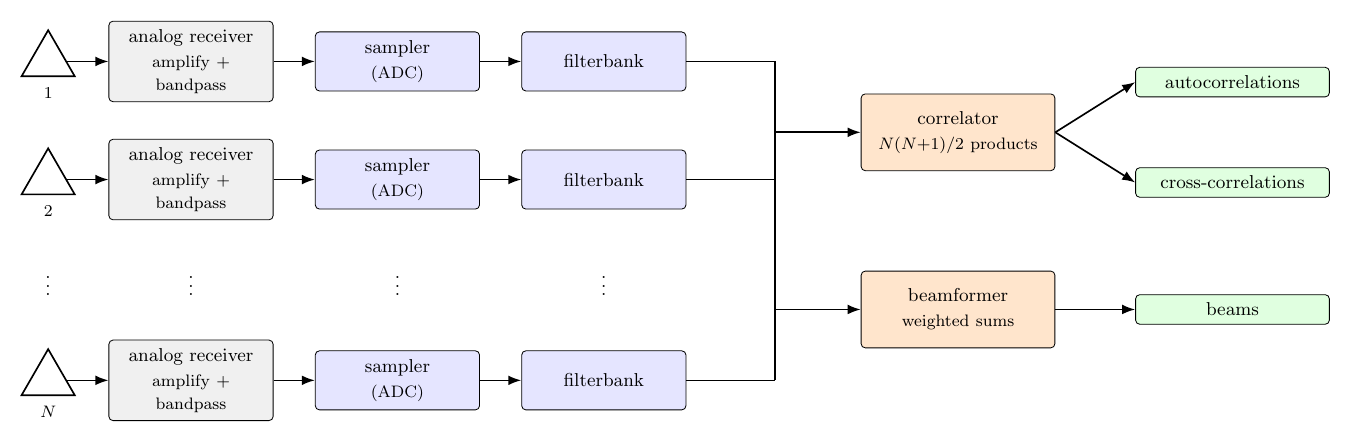}
    \caption{Schematic illustration of a typical ``FX'' radio array signal conditioning and processing system. Signals from each antenna/polarization are amplified and filtered prior to sampling. ``FX'' systems then implement a channelization stage prior to correlation and/or beamforming.}
    \label{fig:1}
\end{figure*}

RFI mitigation in radio-astronomical data relies on data transformations to spaces wherein matched filters can be used to  (\textit{a}) detect RFI over the noise background, and (\textit{b}) isolate RFI from astrophysical signals \citep{an2017,baan2019,buch2025,hugo2026}. In most cases RFI detection is accomplished by applying filters to voltage or power data measured at different times and frequencies; machine-learning approaches \citep[e.g.,][]{dutoit2024} may instead apply data transformations to other bases prior to detection. Typical radio telescope signal processing chains are arranged as shown in Fig.~\ref{fig:1}. The statistical distribution of raw voltage data from individual antennas, immediately after digital samplers (or analog to digital converters / ADCs), is typically determined by the system noise. In cases of strong RFI valid voltage data may be obtained that is RFI dominated. Time-limited RFI that is detected with a significant power relative to the system noise may be identified and excised directly in raw voltage data, or channelized post-filterbank voltage data, using second- or higher-order moment statistics \citep{Buch2023}. Weaker time- and/or frequency-limited RFI is often easily identified in the post-correlation products that represent common radio-telescope data outputs, either in autocorrelations, cross-correlations / visibilities, or autocorrelations of beamformed voltage sums. Several radio astronomy data-analysis packages offer wide selections of methods to identify and excise RFI as non-astronomical correlation data outliers \citep[e.g.,][]{Offringa2010,casa2022,ssins,masoud2023}, or indeed non-astronomical outliers from measured spatial coherence functions \citep[i.e., visibility grids;][]{gridflag}. Techniques have also been developed to detect low-level RFI using coherent combination across radio arrays and spatial filtering \citep{Hellbourg2016}, or correlation with reference antennas \citep{Kesteven2005}.

Radio arrays of immense scale and complexity are now under development, such as LOFAR 2.0 \citep{lofar}, SKA \citep{ska}, the Canadian Hydrogen Observatory and Radio-transient Detector \citep[CHORD;][]{chord}, the Bustling Universe Radio Survey Telescope in Taiwan  \citep[BURSTT;][]{burstt}, the Deep Synoptic Array \citep[DSA;][]{dsa2000}, and the next-generation Very Large Array \citep[ngVLA;][]{ngvla}. Data rates both pre- and post-correlation are rapidly increasing, which together with the need for scientific reproducibility motivate automated RFI excision approaches \citep{Offringa2010}. The move towards advanced, science-ready data products delivered by observatories \citep{dsa2000,srdp} further motivates the real-time detection and excision of RFI \citep{buch2025}. Real-time RFI mitigation offers the additional advantage of being able to apply optimal detection filters prior to data averaging and information loss. However, most studies of RFI mitigation strategies are focused on \textit{how} to identify and mitigate RFI at each stage of signal-processing pipelines (e.g., Fig.~\ref{fig:1}). These strategies, whether automated and/or real-time or neither, are typically limited by the available compute and data throughput. Here I address the parallel question of whether RFI is more easily detectable in the data streams of individual antennas, including autocorrelation data, or in coherent combinations of data in radio arrays (e.g., visibility and beamformed data). This comparative analysis is intended to serve as a guide to the design of new radio-array signal processing systems, in particular for the aforementioned upcoming arrays.

The present work has broader applications beyond the study of naturally occurring astronomical radio sources. Radio arrays are an increasingly important tool in the detection of ultra-high energy cosmic rays, photons, and neutrinos via radio pulses generated by the deflection of air-shower leptons in the Earth's magnetic field \citep{schroder2025}. However, impulsive RFI forms an overwhelming foreground \citep{plant2026}. Radio arrays are increasingly finding uses in passive multistatic radar systems, for example for space situational awareness \citep{msr} and the detection and classification of aerial objects \citep{randall2023}. Radio arrays also realize distinct advantages over individual antennas in the search for extraterrestrial intelligence \citep{ng2021}. In both cases faint signals with little prior information must be separated from RFI. 

I begin this paper with a summary of how radio arrays detect RFI, and the common signal transformations that are used (Section~\ref{sec:model}). I then present an assessment of common RFI excision strategies for radio arrays, with a focus on comparing the detection of RFI in individual-antenna data streams and their incoherent combination, and in coherent data combinations (Section~\ref{sec:strategies}). Simulations are presented of RFI detectability for different emitter scenarios and radio array configurations. This section will demonstrate that RFI detection is usually optimal in the incoherent combination of data from individual antennas. Then, in Section~\ref{sec:autos}, I discuss how this can be achieved in practise, and describe a real-time, automated implementation for the 110-antenna Deep Synoptic Array (DSA-110). The results are discussed in Section~\ref{sec:discussion}, with a focus on practical considerations for upcoming radio arrays, and the conclusions are enumerated in Section~\ref{sec:conc}.

\vspace{0.5cm}

\section{Model for RFI signals and detectability}
\label{sec:model}

In this section I establish the theory of RFI detectability in digital signal processing systems for radio arrays. I begin by defining the formation of visibilities and beams from channelized voltage data. I then derive the factors by which RFI transmissions are suppressed in different estimators due to the individual-element primary beams, near-field defocus and the synthesized-beam response, and fringe tracking. I also discuss RFI detection with and without prior information on the transmitter location. 

\subsection{Voltages, visibilities, and beams}
\label{sec:voltages}

Consider an array of $N$ elements of diameter $D_{\rm a}$ distributed over an extent $D$. Let us assume channelized data, and consider a channel of width $\Delta\nu$ centered on a frequency $\nu$. Element $p$, at position $\mathbf{r}_p$, delivers a complex voltage at time $t$
\begin{equation}
v_p(t,\nu) = g_p\left[\,s_p(t,\nu) + r_p(t,\nu)\,\right] + n_p(t,\nu),
\label{eq:voltage}
\end{equation}
where $g_p$ is the complex gain and $n_p$ is the receiver noise term. The sky term, $s_p$ is given by the  far-field solid-angle integral over the antenna voltage response (primary beam) $A_p$,
\begin{equation}
s_p = \int d\Omega\; A_p(\hat{\mathbf{s}},\nu)\,\mathcal{E}(\hat{\mathbf{s}},t,\nu)\,
e^{-2\pi i\nu\,\hat{\mathbf{s}}\cdot\mathbf{r}_p/c},
\label{eq:sky}
\end{equation}
where $\hat{\mathbf{s}}$ is a direction unit vector, and $\mathcal{E}$ is complex amplitude of the sky electric field, and the RFI term corresponds to a single point emitter at finite distance $R_p = |\mathbf{r}_{\rm R}-\mathbf{r}_p|$ with complex envelope $\tilde a$,
\begin{equation}
r_p = A_p(\hat{\mathbf{s}}_{\rm R},\nu)\,\frac{e^{2\pi i\nu R_p/c}}{R_p}\,\tilde a(t,\nu).
\label{eq:rfi}
\end{equation}
The emitter is characterized by its bandwidth $\Delta\nu_{\rm R}$, its coherence time $t_{\rm R}$ --- the timescale on which $\tilde a$ is stable, characterized by the symbol rate for a digital carrier or the pulse length for a radar --- and its duty cycle $\delta$ within a data integration.

Visibilities are samples of the spatial coherence function, $V_{pq} = \langle v_pv_q^*\rangle$. For a mutually incoherent sky, visibilities are given by the van Cittert--Zernike relation,
\begin{equation}
V_{pq}^{\rm sky} = g_pg_q^*\int d\Omega\; A_pA_q^*\,I(\hat{\mathbf{s}},\nu)\,
e^{-2\pi i\nu\,\hat{\mathbf{s}}\cdot\mathbf{b}_{pq}/c},
\label{eq:vcz}
\end{equation}
where $\mathbf{b}_{pq} = \mathbf{r}_p-\mathbf{r}_q$ and $I$ the apparent sky brightness. Equation~\ref{eq:vcz} holds only for far-field, spatially incoherent signals, and is thus not applicable to the RFI signals of  Eq.~\ref{eq:rfi}. A deterministic emitter at finite distance contributes
\begin{equation}
V_{pq}^{\rm R} = g_pg_q^*A_pA_q^*\,\frac{\langle|\tilde a|^2\rangle}{R_pR_q}\,
e^{2\pi i\nu(R_p-R_q)/c},
\label{eq:vrfi}
\end{equation}
of unit fractional coherence. That is, a single visibility is not intrinsically less sensitive to RFI than a single autocorrelation. All of the suppression derived in Section~\ref{sec:suppression} below arises from averaging Eq.~\ref{eq:vrfi} over
bandwidth, time, or baselines.

A beamformed voltage $b = \sum_p w_pv_p$, where the weights $w_p$ follow $\sum_p|w_p|^2=1$, has power
\begin{equation}
P_{\rm b} = \sum_p |w_p|^2 V_{pp} \;+\; \sum_{p\ne q} w_pw_q^*V_{pq}.
\label{eq:beam}
\end{equation}
A beam is a weighted sum of visibilities together with an autocorrelation term \citep[although for exceptions see]{roy2018}. The first (autocorrelation) term in Eq.~\ref{eq:beam} is independent of the beamformer phases, and cannot be suppressed by any choice of them.

Typically the noise term in Eq.~\ref{eq:voltage}, $n_p$, is defined as a zero-mean Gaussian random variable with $\langle|n_p|^2\rangle = P_{\rm sys} = k_{\rm B}T_{\rm sys}$ (absorbing gain), independent between elements and stronger than the sky terms. For $M = \Delta\nu\,\tau$ independent samples in a time--frequency cell, $\sigma[\hat V_{pp}] = P_{\rm sys}/\sqrt M$ and $\sigma[\hat V_{pq}] = P_{\rm sys}/\sqrt M$. Writing $\mathcal{I} \equiv P_{\rm R}/P_{\rm sys}$ for the per-element interference-to-noise ratio, I define the
reference significance
\begin{equation}
\;\mathcal{R}_1 \equiv \delta\,\mathcal{I}\,\sqrt{\Delta\nu_{\rm R}\,t_{\rm R}}\;
\label{eq:R1}
\end{equation}
as the detection significance of the emitter in the total power of one element, using a filter matched to its time--bandwidth product. Every estimator below is expressed relative to $\mathcal{R}_1$.

\subsection{Suppression of RFI}
\label{sec:suppression}

\subsubsection{Antenna primary beam}

RFI often enters through the far sidelobes of antenna primary-beam responses. Taking the conventional far-sidelobe gain $G_{\rm sl}\simeq1$ (0\,dBi) against a boresight gain $G_0 = \eta_{\rm a}(\pi D_{\rm a}/\lambda)^2$, the power suppression is
\begin{equation}
\mathcal{S}_{\rm PB} \simeq \frac{1}{\eta_{\rm a}}\left(\frac{\lambda}{\pi D_{\rm a}}\right)^{2}.
\label{eq:spb}
\end{equation}
$\mathcal{S}_{\rm PB}$ is common to autocorrelations, visibilities and beams, setting $\mathcal{I}$, and  therefore $\mathcal{R}_1$, but canceling in any ratio of estimators. I absorb it into $\mathcal{I}$ hereafter.

That cancellation presumes the emitter lies in the antenna far field, $R \gg 2D_{\rm a}^2/\lambda$ (the Fraunhofer distance), which is smaller than the corresponding array condition by $(D_{\rm a}/D)^2$. A usual configuration for RFI viewed by a small-element array is to lie simultaneously in the near field of every baseline and the far field of every antenna. RFI emitters within the antenna Fraunhofer distance acquire a quadratic phase error such that the emission fills the sidelobe nulls, and Eq.~\ref{eq:spb} becomes a lower bound and $|A_p|$ acquires a distance dependence that differs between antennas.

\subsubsection{Coherence factor}

I define the per-baseline RFI coherence
\begin{equation}
\gamma_{pq} \equiv \frac{|\langle r_pr_q^*\rangle|}
{\left(P_{\rm R}^{(p)}P_{\rm R}^{(q)}\right)^{1/2}} \;\in\;[0,1],
\label{eq:gamma}
\end{equation}
as the fraction of the RFI power surviving into the visibility, with $\gamma_{pp}\equiv1$ by construction. This can usually be factorized as 
\begin{equation}
\gamma_{pq} \simeq \gamma^{\rm bw}_{pq}\;\gamma^{\rm fr}_{pq}\;\gamma^{\rm sp}_{pq},
\label{eq:gammafac}
\end{equation}
where $\gamma^{\rm bw}$ is bandwidth (delay) smearing across the channel, $\gamma^{\rm fr}$ is fringe-rate time-smearing across the integration, and $\gamma^{\rm sp}$ is the intrinsic spatial coherence of the RFI, reduced by multipath and by multiple independent emitters. I  set $\gamma^{\rm sp}=1$ throughout. I note that $\gamma$ depends on the averaging, and is not a
property of cross-correlation: at short $\tau$ and narrow $\Delta\nu$ it returns to
unity. 

\subsubsection{Bandwidth smearing and near-field defocus}

A deterministic emitter produces a fixed phase on a given baseline, and a fixed phase does not suppress $|V_{pq}|$. Curvature in the radiation field at an array, caused by a near-field emitter, therefore suppresses a single visibility only through the delay across a frequency channel. After far-field delay tracking is applied at the phase center $\hat{\mathbf{s}}_0$, the residual delay has two parts,
\begin{equation}
\Delta\tau_{pq} \approx \underbrace{\frac{(\hat{\mathbf{s}}_0-\hat{\mathbf{s}}_{\rm R})\cdot\mathbf{b}_{pq}}{c}}
_{\rm direction\ offset}
\;+\; \underbrace{\frac{\rho_p^2-\rho_q^2}{2cR}}_{\rm curvature},
\label{eq:dtau}
\end{equation}
with $\rho_p$ the antenna offset transverse to the emitter, and
\begin{equation}
\gamma^{\rm bw}_{pq} = \left|\,\mathrm{sinc}\!\left(\Delta\nu_{\rm eff}\,\Delta\tau_{pq}\right)\right|,
\qquad \Delta\nu_{\rm eff} = \min(\Delta\nu,\Delta\nu_{\rm R}).
\label{eq:gbw}
\end{equation}
The first term of Eq.~\ref{eq:dtau} is usual far-field geometry, and the second is the near-field term (expressed for $R\gg D$), of order $Db_{pq}/2cR$. Their ratio is $\simeq D/[4R\sin(\theta/2)]$ for a boresight angle $\theta$, and thus curvature dominates only for emitters within roughly an array diameter. Because $\Delta\nu_{\rm eff}$ is the emitter bandwidth, Eq.~\ref{eq:gbw} protects better against wideband rather than narrowband RFI.

For tied-array beams the radiation-field curvature is more significant, because beams respond directly to phase. For a uniformly filled circular aperture phased to infinity,
\begin{equation}
\mathcal{S}^{\rm nf}_{\rm coh} = \frac{4\sin^2(U/2)}{U^2},
\qquad U = \frac{\pi D^2}{4\lambda R},
\label{eq:strehl}
\end{equation}
which tends to $4/U^2$ for $U\gg1$. The comparison with Eq.~\ref{eq:gbw} is in the $\nu/\Delta\nu$ term. The beam response nonetheless has a floor, because the autocorrelation term of Eq.~\ref{eq:beam} is phase-independent:
\begin{equation}
\mathcal{S}_{\rm beam} \simeq \max\!\left(N\,\mathcal{S}^{\rm nf}_{\rm coh},\;1\right),
\label{eq:sbeam}
\end{equation}
in units of the single-element response. A coherent beam, with autocorrelations preserved, cannot suppress an emitter below the single-element total-power level. Equation~\ref{eq:sbeam} is the mean over the primary beam. With filling factor $f = N(D_{\rm a}/D)^2$ there are $N/f$ synthesized resolution elements within it, and the expected worst-case coupling is $\ln(N/f) = 2\ln(D/D_{\rm a})$.

\subsubsection{Fringe tracking}

A stationary terrestrial emitter has constant geometric delay in the array frame while the correlator
applies a delay that constantly tracks a phase center fixed on the celestial sphere. RFI emissions therefore acquire the full fringe-stopping phase, at a rate independent of its position on the ground \citep{Athreya2009},
\begin{equation}
f_{\rm F} = \omega_\oplus\,u_\lambda\cos\delta_0
= 0.341\,{\rm Hz}\left(\frac{b_{\rm EW}}{1\,{\rm km}}\right)
\left(\frac{\nu}{1.4\,{\rm GHz}}\right)\cos\delta_0,
\label{eq:ffringe}
\end{equation}
where $\omega_\oplus$ is the Earth rotation rate, $u_\lambda$ ($b_{\rm EW}$) is the east-west projected baseline in wavelengths (km), and $\delta_0$ is the declination of the phase center. Averaging over $\tau$ gives $\gamma^{\rm fr} = |\mathrm{sinc}(f_{\rm F}\tau)|$. The suppression is bound by the envelope $1/(\pi f_{\rm F}\tau)$. Additionally, real emitters may be amplitude modulated, spreading the fringe into sidebands.

For RFI detection the matched choice is $\tau = t_{\rm R}$, since integrating beyond the emitter's coherence time adds noise without adding signal. Baselines shorter than
\begin{equation}
b_{\rm crit} = \frac{\lambda}{\omega_\oplus t_{\rm R}\cos\delta_0}
= 2.9\,{\rm km}\left(\frac{\lambda}{0.214\,{\rm m}}\right)
\left(\frac{1\,{\rm s}}{t_{\rm R}}\right)\sec\delta_0
\label{eq:bcrit}
\end{equation}
have no fringe-rate suppression at all. The scaling with $t_{\rm R}$ is strong: $b_{\rm crit} = 2900$\,km for $t_{\rm R} = 1$\,ms, 2.9\,km for 1\,s and 290\,m for 10\,s. Fringe tracking provides no suppression of interference that is impulsive on sub-second timescales, in any ground-based array. It is effective only against persistent emitters observed with long integrations, which is the regime in which they would in any case be detected easily in total power. A corollary is that non-sidereal apparent motion --- aircraft, low-Earth-orbit satellites --- lowers $\gamma^{\rm fr}$ independent of baseline length.

\subsection{Coherent and incoherent RFI detection}
\label{sec:detection}

\subsubsection{Estimators and trials}

A statistic that selects a maximum over $N_{\rm trials}$ approximately independent tests must be detected above a higher threshold for a fixed false-alarm rate. To first order,
\begin{equation}
\rho_{\rm eff} \simeq \left(\rho_0^2 + 2\ln N_{\rm trials}\right)^{1/2}.
\label{eq:trials}
\end{equation}
I therefore compare estimators by their detection margin
\begin{equation}
\mathcal{M} \equiv \mathcal{R}/\rho_{\rm eff},
\label{eq:margin}
\end{equation}
taking $\rho_0 = 5$, so that $\mathcal{M}>1$ implies detection. Every statistic scans the available
time--frequency cells; those that additionally scan baselines, $uv$ cells or trial emitter positions incur a false-alarm-rate penalty.

\begin{table}[!t]
\centering
\caption{Symbols used throughout.}
\label{tab:symbols}
\begin{tabular}{cc}
\toprule
Symbol & Meaning \\
\midrule
$N$, $N_{\rm b}$ & number of antennas; number of baselines, $N(N-1)/2$ \\
$D$, $D_{\rm a}$ & array extent; element diameter \\
$f$ & array filling factor, $N(D_{\rm a}/D)^2$ \\
$\Delta\nu$, $\tau$ & channel width; integration time \\
$M$ & samples per cell, $\Delta\nu\,\tau$ \\
$\Delta\nu_{\rm R}$, $t_{\rm R}$, $\delta$ & emitter bandwidth, coherence time, duty cycle \\
$\theta$ & boresight angle between pointing and emitter \\
$R$ & emitter range \\
$\mathcal{I}$ & per-element interference-to-noise ratio, $P_{\rm R}/P_{\rm sys}$ \\
$\mathcal{R}_1$ & reference significance, Eq.~\ref{eq:R1} \\
$\mathcal{S}_{\rm PB}$ & far-sidelobe power suppression, Eq.~\ref{eq:spb} \\
$\gamma_{pq}$ & per-baseline RFI coherence, Eq.~\ref{eq:gamma} \\
$\gamma^{\rm bw}$, $\gamma^{\rm fr}$, $\gamma^{\rm sp}$ & bandwidth, fringe, and spatial factors of $\gamma$ \\
$\langle\gamma\rangle$, $\bar\gamma_{\rm rms}$ & mean and rms of $\gamma_{pq}$ over baselines \\
$d$ & time-differencing factor, Eq.~\ref{eq:ddiff} \\
$f_{\rm F}$, $b_{\rm crit}$ & fringe rate; critical baseline, Eqs.~\ref{eq:ffringe}, \ref{eq:bcrit} \\
$\mathcal{S}^{\rm nf}_{\rm coh}$ & near-field defocus of a coherent beam, Eq.~\ref{eq:strehl} \\
$\rho_0$, $\rho_{\rm eff}$ & single-trial and trials-corrected thresholds, Eq.~\ref{eq:trials} \\
$\mathcal{M}$ & detection margin, $\mathcal{R}/\rho_{\rm eff}$ \\
\bottomrule
\end{tabular}
\end{table}
\vspace{0.5cm}

The estimators of interest, with their significances relative to $\mathcal{R}_1$, are
\begin{equation}
\mathcal{R}_{\rm auto} = \sqrt{N}\,\mathcal{R}_1
\label{eq:Rauto}
\end{equation}
for the sum of the $N$ autocorrelations, in which the emitter appears at unit coherence and linearly in
power;
\begin{equation}
\mathcal{R}_{pq} = \sqrt2\,\gamma_{pq}\,\mathcal{R}_1
\label{eq:Rbl}
\end{equation}
for a single baseline, and
\begin{equation}
\mathcal{R}_{\rm coh} = \sqrt{2}\left(\textstyle\sum_{p<q}\gamma_{pq}^2\right)^{1/2}\mathcal{R}_1
\simeq \bar\gamma_{\rm rms}\,N\,\mathcal{R}_1
\label{eq:Rcoh}
\end{equation}
for a coherent sum over all baselines where the phase of the RFI emitter is known, where
$\bar\gamma_{\rm rms}\equiv\langle\gamma^2\rangle^{1/2}$.

\subsubsection{Detection without phase knowledge}

Equation~\ref{eq:Rcoh} recovers the sensitivity scaling with $N$, but requires that the visibility phase $\Phi_{pq} = 2\pi\nu(R_p-R_q)/c$ be known. The emitter position is in general unknown, so the phase must either be supplied externally, or scanned over a grid of trial positions $\mathbf{x}$. Writing $\hat{\mathbf{V}}$ for the $N\times N$ matrix of measured visibilities $\hat V_{pq}$, and $\mathbf{w}(\mathbf{x})$ for the near-field visibility steering vector with $w_p \propto e^{-2\pi i\nu R_p(\mathbf{x})/c}$, each trial evaluates the off-diagonal part of the quadratic form $T(\mathbf{x}) = \mathbf{w}^\dagger\hat{\mathbf{V}}\mathbf{w}$. The number of trials is set by the number of resolution elements in the search volume: $(D/D_{\rm a})^2 = N/f$ angular cells within the primary beam, and $U_{\max} = \pi D^2/(4\lambda R_{\min})$ cells in range (see Eq.~\ref{eq:strehl}), where $R_{\min}$ is the smallest emitter distance searched. Taking $R_{\min}$ to be the antenna Fraunhofer distance $2D_{\rm a}^2/\lambda$, inside which Eq.~\ref{eq:rfi} no longer holds, we have $U_{\max} = (\pi/8)(N/f)$ and $N_{\rm trials} \simeq (\pi/8)(N/f)^2$. Equation~\ref{eq:trials} then yields a threshold trials penalty $\rho_{\rm eff}/\rho_0 \lesssim 1.6$ for $N/f \lesssim 10^4$, against a potential gain of $\bar\gamma_{\rm rms}\sqrt{N}$ (Eqs.~\ref{eq:Rauto} and \ref{eq:Rcoh}). A search over arbitrary spatial modes, with $\mathbf{w}$ ranging over the unit sphere in $\mathbb{C}^N$, is quite different: the covering number of that sphere gives $\ln N_{\rm trials}\sim cN$ and a penalty $\sim\sqrt{N}$, which cancels the coherent gain exactly. The distinction lies in the dimensionality of the search, not in the form of the statistic, which is linear in  $\hat{\mathbf{V}}$ in both cases.

The alternative to a search over emitter positions is an incoherent statistic. The most sensitive in use is that of \citet{ssins}, which differences visibilities between adjacent integrations to remove the sky and then averages $|\Delta V_{pq}|$ over baselines, discarding phase. Two properties govern its performance. First, the differencing doubles the noise variance while multiplying a persistent emitter by $|1-e^{2\pi if_{\rm F}\Delta t}|$, for a net factor
\begin{equation}
d = \sqrt2\left|\sin\left(\pi f_{\rm F}\Delta t\right)\right|,
\label{eq:ddiff}
\end{equation}
limited to unity once adjacent samples are uncorrelated. Equation~\ref{eq:ddiff} vanishes for slowly fringing RFI, so the compact configurations that best preserve $\gamma$ are also those whose fringes are too slow to survive differencing. Second, the magnitude operation is a square-law detector: for $|V^{\rm R}|\ll\sigma$ the Rice mean expands as $\mathbb{E}|\hat V|\simeq\sigma\sqrt{\pi/2}\,(1+|V^{\rm R}|^2/4\sigma^2)$, so the response is quadratic in the RFI amplitude and the array gain collapses from $N$ to $\sqrt{N}$,
\begin{equation}
\mathcal{R}_{\rm incoh} \simeq \bigl(\bar\gamma d\,\mathcal{R}_1\bigr)^2\sqrt{N_{\rm b}},
\label{eq:Rincoh}
\end{equation}
with $N_{\rm b} = N(N-1)/2$. Above the per-baseline noise the behavior is quite different:
$|\Delta V_{pq}|\simeq|V^{\rm R}|+\mathrm{Re}(\epsilon_{pq}e^{-i\Phi_{pq}})$, so each term is the coherent projection onto the phase the emitter itself defines, and Eq.~\ref{eq:Rincoh} reverts to Eq.~\ref{eq:Rcoh} up to the Cauchy--Schwarz factor $\langle\gamma\rangle/\bar\gamma_{\rm rms}\le1$. The
$\sqrt{N_{\rm b}}$ sensitivity scaling is always a coherent gain; a phase-unknown statistic obtains it only when the emitter is bright enough to supply its own phase reference.

One further caveat applies to any statistic that involves averaging baseline amplitudes. Distinct visibilities are uncorrelated, but their amplitudes are not, because $|\hat V_{pq}|$ inherits the sample power of the shared element $p$. For $Y = \langle|\hat V_{pq}|\rangle$,
\begin{equation}
\frac{\mathrm{Var}\,Y}{\langle Y\rangle^2} = \frac{4/\pi-1}{N_{\rm b}} + \frac{1}{NM},
\label{eq:sharedvar}
\end{equation}
the two terms crossing at $M = 1.83(N-1)$. Below that the $\sqrt{N_{\rm b}}$ advantage is lost and the precision reverts to that of the autocorrelation sum.

\subsubsection{Section summary}

Combining Eqs.~\ref{eq:margin}--\ref{eq:Rincoh}, and writing $\gamma d$ for the effective coherence of the phase-unknown statistic, the comparison that Section~\ref{sec:strategies} below evaluates numerically reduces above the per-baseline noise to
\begin{equation}
\;\frac{\mathcal{M}_{\rm incoh}}{\mathcal{M}_{\rm auto}} = \langle\gamma d\rangle\sqrt{N-1},
\qquad
\frac{\mathcal{M}_{\rm coh}}{\mathcal{M}_{\rm auto}} = \bar\gamma_{\rm rms}\sqrt{N-1}\,
\frac{\rho_{\rm auto}}{\rho_{\rm coh}}\;
\label{eq:key}
\end{equation}
and below it acquires a further factor $\mathcal{O}(\bar\gamma d\,\mathcal{R}_1)$ from the loss of a square-law detector in the weak signal regime. The advantage of a cross-correlation statistic over the autocorrelation sum is the RFI coherence multiplied the square root of the number of elements. The crossover follows immediately:
\begin{equation}
\langle\gamma d\rangle > (N-1)^{-1/2},
\label{eq:crossover}
\end{equation}
which is $0.20$ for a 27-element array, $0.13$ for 64 and $0.025$ for 1650. A large array needs less
coherence to make cross-correlation detection worthwhile, but a large array also destroys coherence. Which effect wins is not decidable in general, and Section~\ref{sec:strategies} addresses this case by case.

\section{Assessment of RFI excision strategies for radio arrays}
\label{sec:strategies}

Section~\ref{sec:model} reduced the comparison between antenna-based and array-based RFI detection to the per-baseline coherence $\gamma_{pq}$ and the number of antennas $N$. Both depend on the array configuration, the emitter, and the choices of channel width and integration time, and $\gamma_{pq}$ varies by orders of magnitude between baselines of the same array. In this section I evaluate Eq.~\ref{eq:key} numerically for four array configurations and five emitters.

\subsection{Simulation method}
\label{sec:method}

For each combination of array, emitter and pointing I compute:
\begin{enumerate}
    \item the exact geometric delay from the emitter to every element, with the the far-field delay applied while tracking the phase center subtracted, giving the residual delay of Eq.~\ref{eq:dtau} on every baseline;
    \item its time derivative, by differencing over the integration with both Earth rotation and emitter motion included;
    \item the coherence $\gamma_{pq}$ from Eqs.~\ref{eq:gbw} and \ref{eq:ffringe};  
    \item the differencing factor $d$ of Eq.~\ref{eq:ddiff}; and 
    \item the detection margin $\mathcal{M}=\mathcal{R}/\rho_{\rm eff}$ for each estimator.
\end{enumerate}
Calculations are done for pointings along a great circle from the emitter direction through the zenith so that the boresight angle $\theta$ runs from $0^\circ$ to $\sim\!120^\circ$ with the antenna always above $10^\circ$ elevation. Results are computed  per channel per second. The channel bandwidths for each array under consideration are given in Table~\ref{tab:arrays}. The results below, in particular for suppression due to emitter fringe rates, are dependent on the 1\,s integration, although this is to an order of magnitude standard for the arrays under consideration (after the application of sidereal fringe tracking). 

The per-antenna interference-to-noise ratio $\mathcal{I}$ is set by hand. Fixing $\mathcal{I}$ removes the primary beam from the comparison, since from Section~\ref{sec:suppression} it multiplies autocorrelations and cross-correlations identically and cancels in every ratio, and it isolates the question of how the geometry of an emitter determines where it is best detected. Two levels are used, chosen to bracket the transition between the signal to noise regimes of Eq.~\ref{eq:Rincoh}.

\emph{Case~L} is the front-end linearity limit. For a $b$-bit sampler with the system noise set to $\sigma_{\rm c}$ counts rms and $k$ standard deviations of clipping headroom, $\mathcal{I}_{\rm lin} = (2^{\,b-1}/k\sigma_{\rm c})^2-1$, which for $b=8$, $\sigma_{\rm c}=10$ and $k=4$ gives $9.2$. I adopt $\mathcal{I}=10$. Above this the front end compresses and the model of Eq.~\ref{eq:voltage} fails.

\emph{Case~F} is $\mathcal{I}=10^{-3}$, four orders of magnitude fainter. This places the emitter at or below the noise of one element in one channel in one second for every array considered ($\mathcal{R}_1 = 1.00$ for the VLA, $0.46$ for MeerKAT, $0.36$ for DSA), and far below the noise of any single baseline. This is the regime of importance for deep continuum and spectral-line observations.

I note that gain and bandpass stability are neglected throughout, so the autocorrelation statistics are quoted at their
radiometric limits. This assumption favors detection in the autocorrelations. 

\subsection{Array and RFI emitter configurations}
\label{sec:configs}

Four arrays are used (Table~\ref{tab:arrays}), spanning physical extent at roughly fixed collecting
area (VLA-D versus VLA-A), element count at fixed extent (VLA-D versus MeerKAT), and both together (DSA). The DSA configuration used is representative, and is not the final design. 

Five RFI emitters are used (Table~\ref{tab:emitters}), spanning distances from 5\,km to geosynchronous orbits and transverse velocity from zero to $7.6$\,km\,s$^{-1}$. Only their geometry and signal structure are relevant: distance, elevation, velocity, carrier frequency, occupied bandwidth, and duty cycle. Their powers are set by Section~\ref{sec:method}, so no propagation or link-budget model is required.

\begin{table}[!t]
\centering
\caption{Array configurations. $\tilde b$ is the median baseline. The last three columns are the $5\sigma$-equivalent
trials-corrected thresholds of Eq.~\ref{eq:trials}.}
\label{tab:arrays}
\begin{tabular}{lrrrrrrrr}
\toprule
& $N$ & $D_{\rm a}$ & $b_{\max}$ & $\tilde b$ & $T_{\rm sys}$ & $\Delta\nu$ &
\multicolumn{2}{c}{$\rho_{\rm eff}$} \\
\cmidrule(lr){8-9}
& & (m) & (km) & (km) & (K) & (kHz) & auto & coh. \\
\midrule
VLA-D   & 27   & 25.0 & 1.04  & 0.38  & 26 & 1000 & 7.18 & 8.05 \\
MeerKAT & 64   & 13.5 & 7.70  & 0.96  & 20 & 209  & 7.38 & 8.43 \\
DSA     & 1650 & 5.0  & 20.30 & 7.94  & 25 & 130  & 7.50 & 9.27 \\
VLA-A   & 27   & 25.0 & 36.37 & 13.27 & 26 & 1000 & 7.18 & 8.05 \\
\bottomrule
\end{tabular}
\end{table}
\vspace{0.5cm}

\begin{table}[!h]
\centering
\caption{RFI emitter geometry and signal structure. $v$ is the transverse velocity and $\delta$ the duty cycle
within a 1\,s integration.}
\label{tab:emitters}
\begin{tabular}{lrrrrrr}
\toprule
& $\nu$ & $R$ & el. & $v$ & $\Delta\nu_{\rm R}$ & $\delta$ \\
& (GHz) & & ($^\circ$) & (m\,s$^{-1}$) & (MHz) & \\
\midrule
Handset & 1.75  & 5\,km       & 0  & 0    & 10    & 1 \\
Tower   & 1.85  & 100\,km     & 0  & 0    & 10    & 1 \\
ADS-B   & 1.09  & 14.1\,km    & 45 & 250  & 2.6   & $2.4\times10^{-4}$ \\
GNSS    & 1.575 & 20\,200\,km & 45 & 3900 & 2.046 & 1 \\
D2C     & 1.99  & 550\,km     & 45 & 7600 & 5     & 1 \\
\bottomrule
\end{tabular}
\end{table}

\subsection{RFI coherence}
\label{sec:coherence}

\begin{figure}
\centering
\includegraphics[width=0.9\columnwidth]{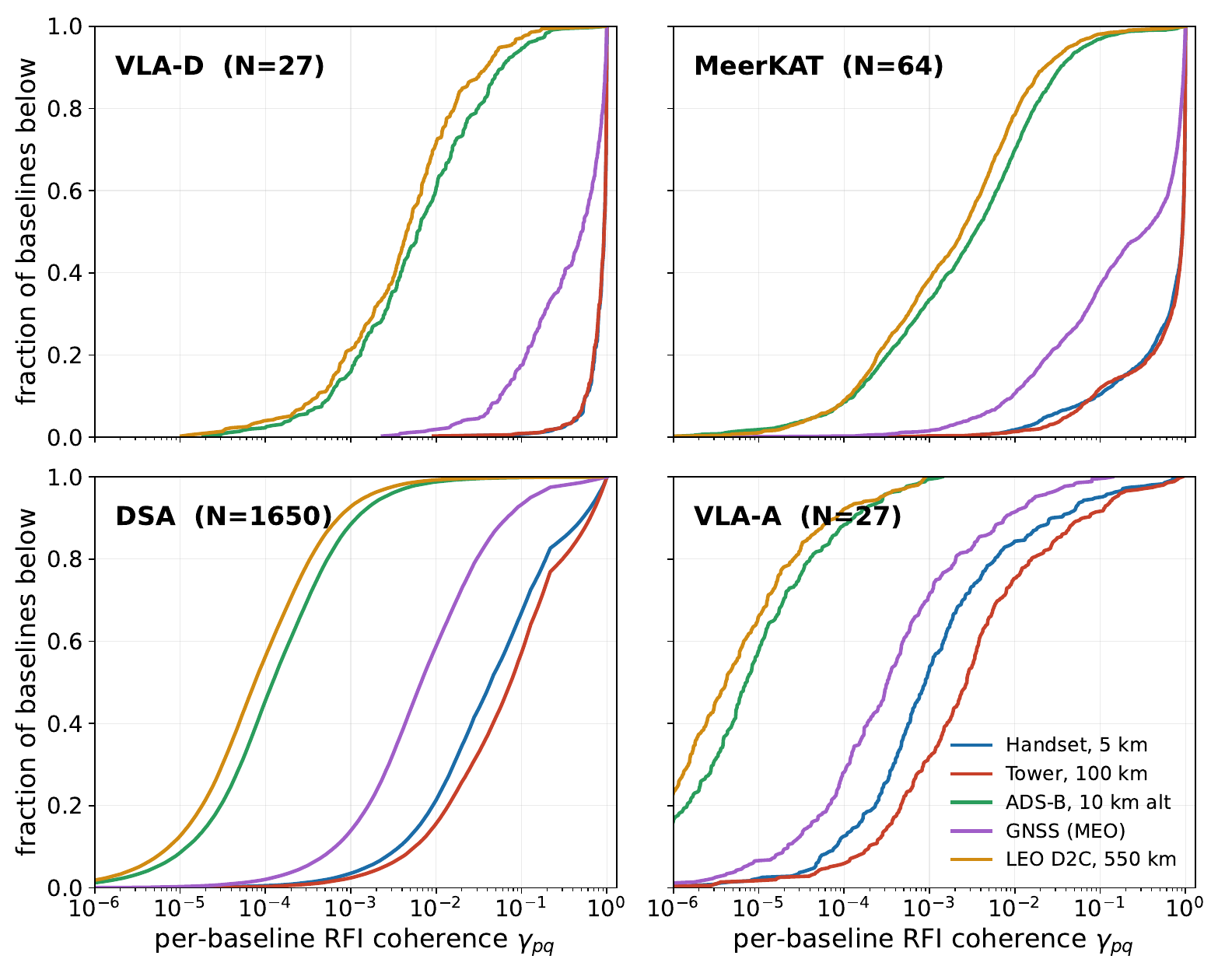}
\caption{Cumulative distribution of the per-baseline RFI coherence $\gamma_{pq}$ for each array and RFI emitter, at $\theta=45^\circ$ in a 1\,s integration and one channel. A value of unity means the visibility sees the RFI at full strength.}
\label{fig:gamma}
\end{figure}

Table~\ref{tab:gamma} decomposes the residual delay and gives the resulting coherence for the median
baseline of each array. Fig.~\ref{fig:gamma} shows the full distributions. Three results are noteworthy.

First, compact arrays provide minimal suppression of a stationary emitter. For the VLA-D configuration, the median residual delay is $0.2\,\mu$s and the median fringe rate $0.06$\,Hz, so over a 1\,MHz channel and a 1\,s integration both sinc-function arguments in Eqs.~\ref{eq:gbw} and \ref{eq:ffringe} are far below unity and $\gamma\simeq0.93$. The handset and the tower are as visible in a VLA-D visibility as in an autocorrelation. MeerKAT behaves the same way: half its baselines are shorter than 1\,km and $\gamma^{\rm bw}=0.97$.

Second, emitter motion rather than array size is what usually suppresses coherence. ADS-B and the direct to cell satellite have $\gamma^{\rm fr}\sim10^{-2}$ even for VLA-D and MeerKAT, because for $250$\,m\,s$^{-1}$ and $7.6$\,km\,s$^{-1}$ respectively the geometric delay changes fast enough to rotate the fringe many times per second. A compact array offers no protection against a moving emitter.

Third, near-field curvature is usually a sub-dominant term. The two delay columns of Table~\ref{tab:gamma} separate far-field geometry from the curvature term of Eq.~\ref{eq:dtau}. Curvature exceeds the direction-offset term in only two cases, both the 5\,km handset, and contributes $10\%$ for the 100\,km tower. Near-field treatment of visibilities is primarily needed for interference on the scale of array diameters.

\subsection{Estimators}
\label{sec:estimators}

\begin{figure*}
\centering
\includegraphics[width=0.86\textwidth]{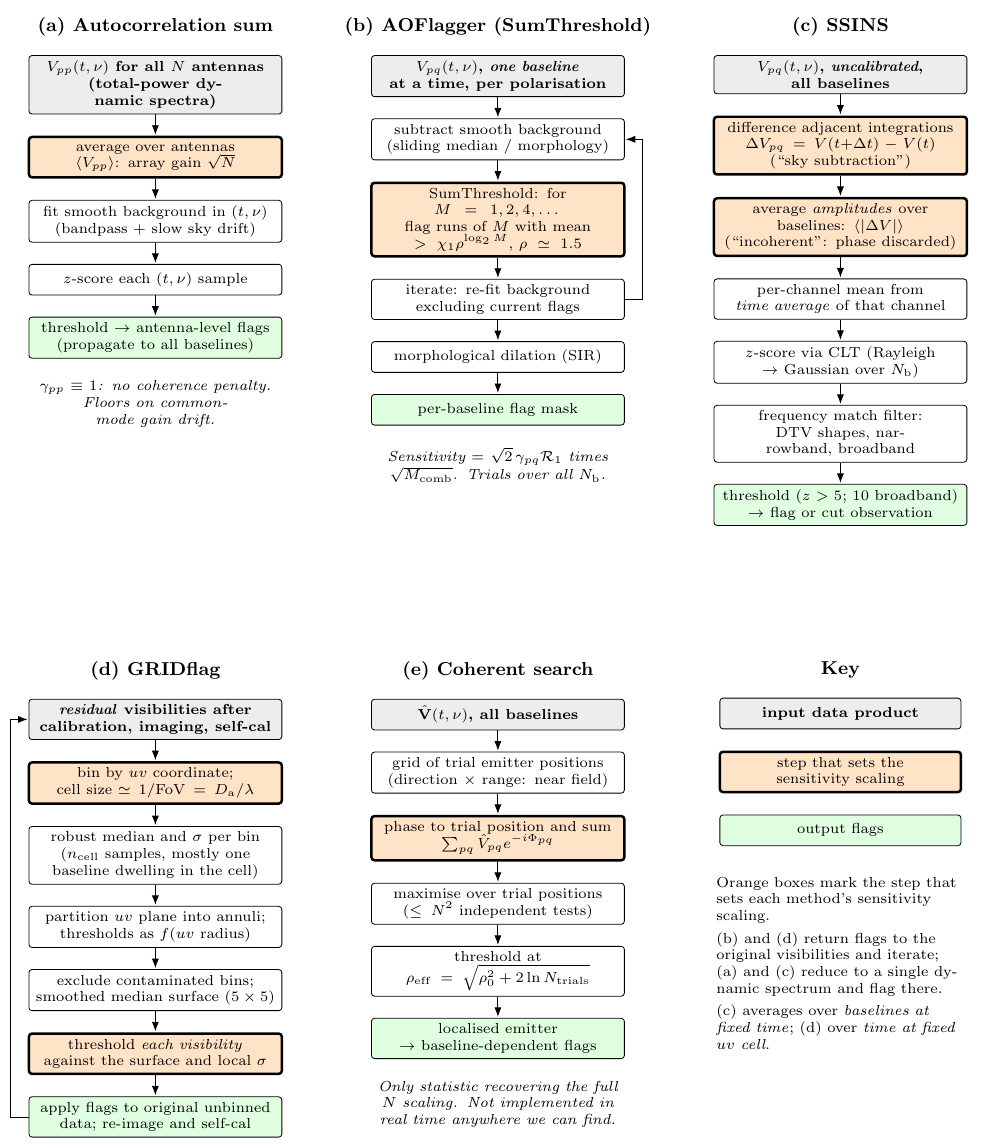}
\caption{Processing chains of the estimators compared herein. Orange marks the step that sets each method's
sensitivity scaling: array averaging for (a), combinatorial thresholding for (b), differencing and baseline averaging for (c), $uv$ gridding for (d),
 and coherent phasing for (e). (a) and (d) reduce the array to a single dynamic spectrum and flag within it, whereas (b) and (c)
return flags to the visibilities and iterate; and that (d) averages over baselines at fixed time while (c)
averages over time at fixed $uv$ cell.}
\label{fig:algorithms}
\end{figure*}

Five statistics are computed, corresponding to RFI detection methods in use. Each signal processing chain is sketched in Fig.~\ref{fig:algorithms}. The ``median baseline'' is the significance on a typical baseline, plotted to show the spread. ``AOFlagger'' represents the SumThreshold algorithm applied to each baseline's time--frequency plane independently \citep{Offringa2010}, modeled as the best baseline times a combinatorial gain $\sqrt{M_{\rm comb}}$ with $M_{\rm comb}=8$. Some baseline in any array has a small residual delay and a slow fringe, but AOFlagger tests all $N_{\rm b}$ baselines and has the corresponding trials factor. ``GRIDflag'' compares visibilities sharing a $uv$ cell \citep{gridflag}, gaining $\sqrt{n_{\rm eff}}$ over a single baseline, where $n_{\rm eff}$ is the number of simultaneous visibilities sharing a cell through the gridding convolution function. ``Incoherent cross'' is the incoherent baseline-amplitude statistic of \citet{ssins}, computed with the exact Rice mean and variance so that it is correct in both regimes of Eq.~\ref{eq:Rincoh}, and including the differencing factor $d$ and the shared-element variance term of Eq.~\ref{eq:sharedvar}. ``Coherent (searched)'' phases all baselines to a trial RFI emitter position and takes the maximum over a grid, with a trials-factor penalty described in Section~\ref{sec:detection}.

\subsection{Results}
\label{sec:results}

\begin{figure*}
\centering
\includegraphics[width=\textwidth]{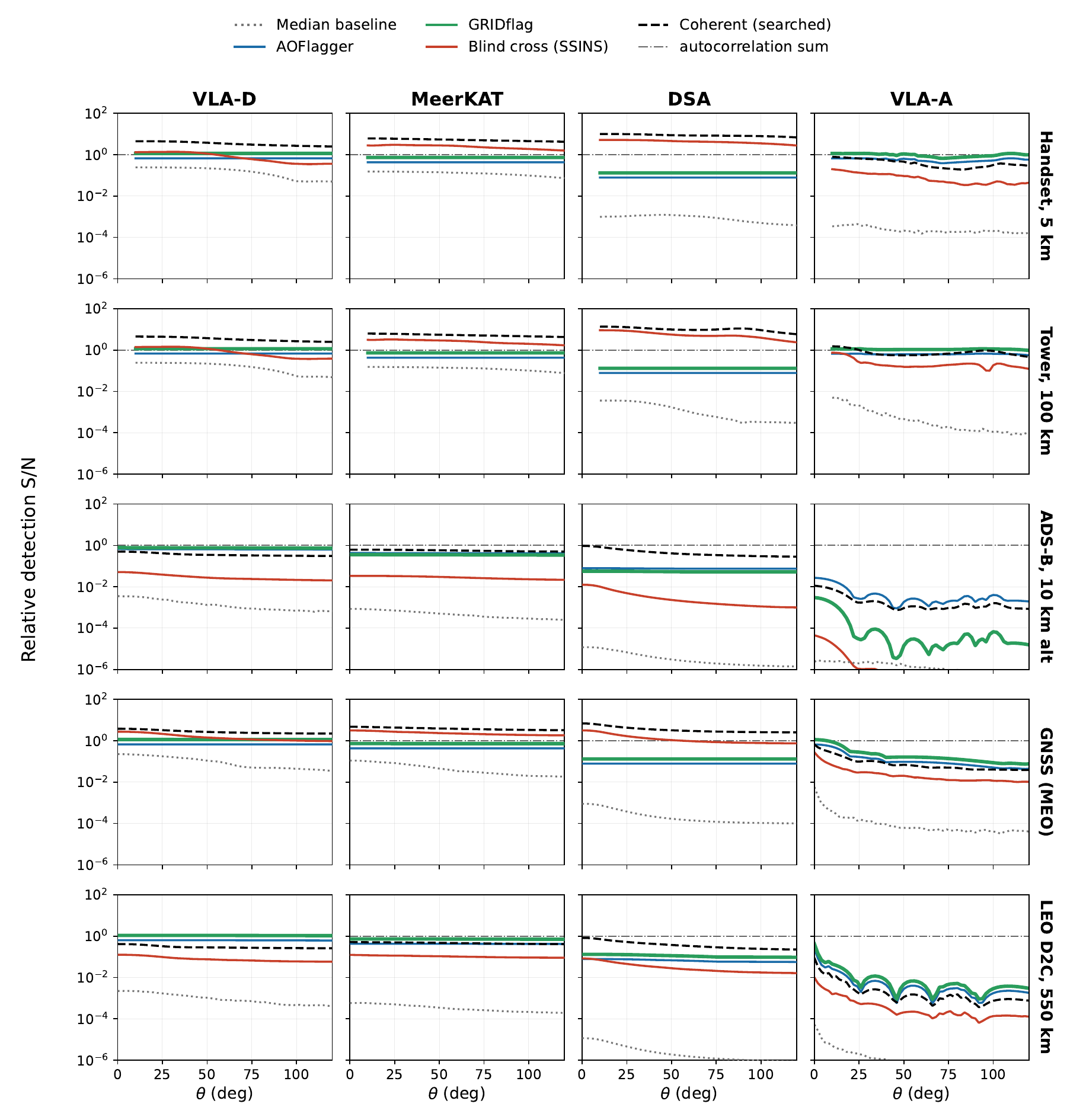}
\caption{Case~L ($\mathcal{I}=10$, the front-end linearity limit). Detection margin of each
cross-correlation statistic relative to the autocorrelation sum, per channel per second, with trials factors
folded in. Above the dot-dashed line the cross-correlations detect the RFI more securely than the sum of
the autocorrelations.}
\label{fig:caseL}
\end{figure*}

\begin{figure*}
\centering
\includegraphics[width=\textwidth]{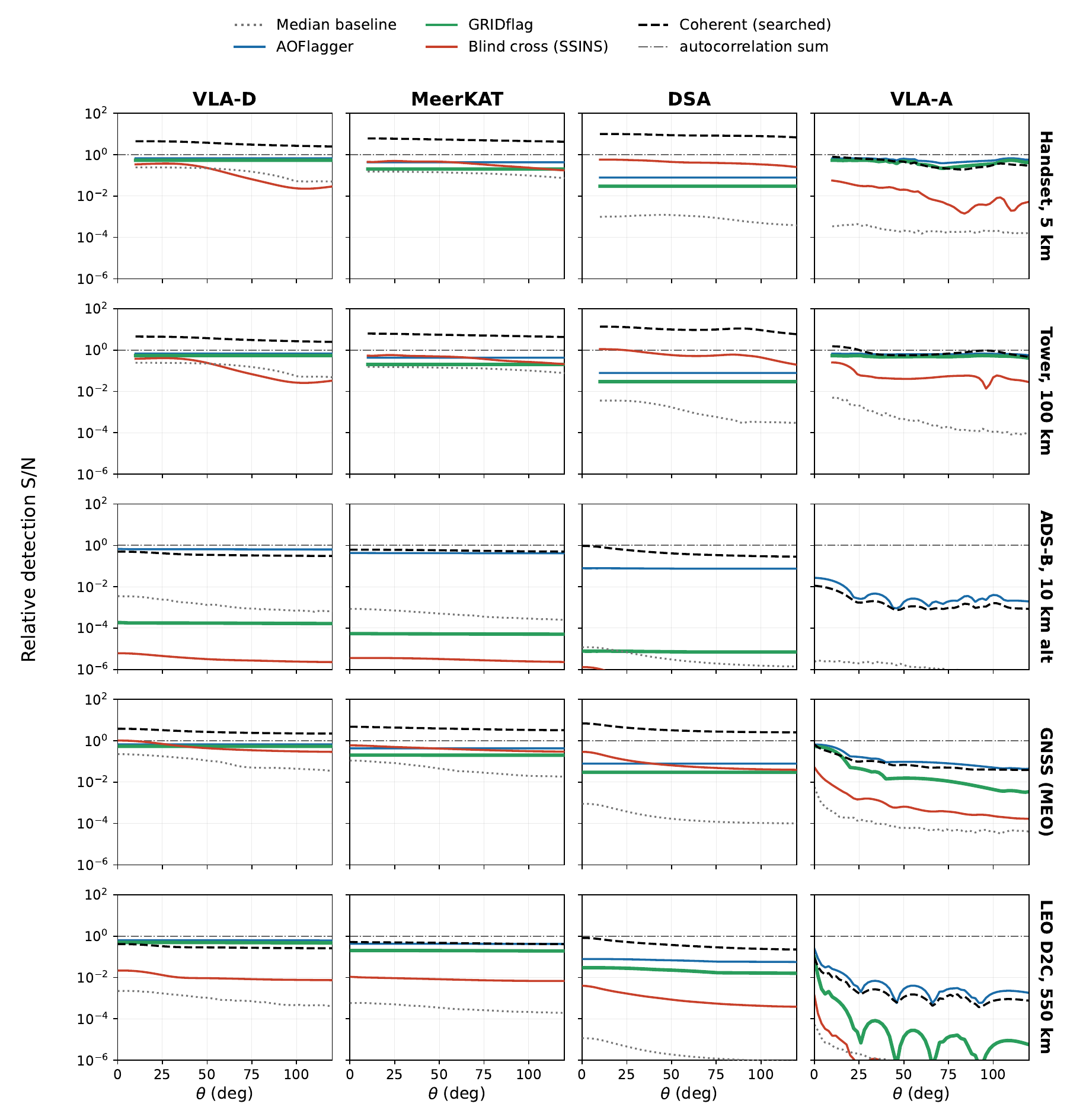}
\caption{Case~F ($\mathcal{I}=10^{-3}$, at or below the single-element noise). Legend as in Fig.~\ref{fig:caseL}. The
coherent curve is unchanged, being linear in $\mathcal{R}_1$, while the phase-unknown curve has dropped by one
to four orders of magnitude, which is the square-law loss of Eq.~\ref{eq:Rincoh}.}
\label{fig:caseF}
\end{figure*}

Figures~\ref{fig:caseL} and \ref{fig:caseF} give the detection margin of each statistic relative to the
autocorrelation sum, for the two power levels. The ratios are nearly independent of boresight angle $\theta$. The primary beam multiplies the RFI emission into autocorrelations and cross-correlations identically, and cancels in any ratio, and the remaining residual $\theta$-dependence comes from the
changing residual delays as the phase center moves.

 \begin{table}
\centering
\caption{Delay on the median baseline at $\theta=45^\circ$, split into the far-field
direction-offset and near-field curvature terms of Eq.~\ref{eq:dtau}, and the resulting coherence.
$\langle\gamma d\rangle$ and $(\gamma d)_{\rm rms}$ are the mean and rms over all baselines of the product of
Eqs.~\ref{eq:gamma} and \ref{eq:ddiff}.}
\label{tab:gamma}
\begin{tabular}{llrrrrr}
\toprule
& & \multicolumn{2}{c}{$|\Delta\tau|$ ($\mu$s)} & $f_{\rm F}$ & & \\
\cmidrule(lr){3-4}
Array & Emitter & dir. & curv. & (Hz) & $\langle\gamma d\rangle$ & $(\gamma d)_{\rm rms}$ \\
\midrule
VLA-D & Handset & 0.20 & 0.01 & 0.057 & 0.243 & 0.293 \\
VLA-D & Tower   & 0.20 & 0.00 & 0.060 & 0.252 & 0.302 \\
VLA-D & ADS-B   & 0.49 & 0.01 & 12.5  & 0.023 & 0.064 \\
VLA-D & GNSS    & 0.49 & 0.00 & 0.205 & 0.299 & 0.390 \\
VLA-D & D2C     & 0.49 & 0.00 & 17.6  & 0.015 & 0.053 \\
\midrule
MeerKAT & Handset & 0.53 & 0.08 & 0.151 & 0.358 & 0.443 \\
MeerKAT & Tower   & 0.53 & 0.00 & 0.159 & 0.375 & 0.460 \\
MeerKAT & ADS-B   & 1.29 & 0.06 & 31.2  & 0.018 & 0.075 \\
MeerKAT & GNSS    & 1.29 & 0.00 & 0.523 & 0.305 & 0.428 \\
MeerKAT & D2C     & 1.29 & 0.00 & 45.2  & 0.013 & 0.059 \\
\midrule
DSA & Handset & 3.83 & 6.88 & 1.33 & 0.116 & 0.221 \\
DSA & Tower   & 3.83 & 0.37 & 1.41 & 0.149 & 0.255 \\
DSA & ADS-B   & 9.30 & 2.70 & 296  & 0.001 & 0.011 \\
DSA & GNSS    & 9.30 & 0.00 & 5.31 & 0.028 & 0.085 \\
DSA & D2C     & 9.30 & 0.07 & 471  & 0.001 & 0.010 \\
\midrule
VLA-A & Handset & 7.05 & 17.30 & 2.00 & 0.019 & 0.085 \\
VLA-A & Tower   & 7.05 & 0.76 & 2.11 & 0.031 & 0.113 \\
VLA-A & ADS-B   & 17.03& 9.34 & 375  & $10^{-4}$ & $2\times10^{-4}$ \\
VLA-A & GNSS    & 17.03& 0.01 & 7.18 & 0.004 & 0.013 \\
VLA-A & D2C     & 17.03& 0.28 & 616  & $<10^{-4}$ & $10^{-4}$ \\
\bottomrule
\end{tabular}
\end{table}
\vspace{0.5cm}

Compact arrays favor the cross-correlations, whereas extended sparse arrays do not. In Case~L, for VLA-D and MeerKAT observing a stationary emitter, the incoherent and coherent statistics beat the autocorrelation sum by factors of $1.2$--$5.6$. DSA does better still, $4.8$--$10.3$ for the handset and the tower, because the baseline factor $\sqrt{N-1}=40.6$ more than compensates for its lower coherence. For VLA-A the autocorrelation sum is superior for any emitter.

Moving RFI emitters favor the autocorrelations everywhere. ADS-B and D2C fall below unity in almost every case, including the compact ones, by one to seven orders of magnitude. No array geometry protects a visibility against an emitter for which the delay changes by many wavelengths per second, and correspondingly none makes a visibility a good detector of one.

Below the single-element noise the phase-unknown statistic collapses. Between Case~L and Case~F the coherent curves do not move, since its margin and that of the autocorrelation sum are both linear in $\mathcal{R}_1$, so their ratio is independent of emitter brightness. The incoherent curve falls by one to four orders of magnitude: from $1.24$ to $0.28$ for VLA-D's handset, from $4.76$ to $0.47$ for DSA's, and from $6.07$ to $0.63$ for DSA's tower. In Case~L it exceeds the autocorrelation sum in eight array--emitter combinations, whereas in Case~F it never does so.

\subsection{Detectability in a tied-array beam}
\label{sec:beam}

Tied-array beams formed with arrays, rather than visibilities, are often used for fast time domain studies. Equation~\ref{eq:beam} splits the beam power into an incoherent autocorrelation term and a coherent sum over baselines, and only the second depends on the beamformer phases. I evaluate the coherent response exactly, as
\begin{equation}
\mathcal{S}_{\rm beam} = \frac{1}{N}\left\langle\left|\sum_p
e^{2\pi i\nu\,\Delta\tau_p}\right|^2\right\rangle,
\label{eq:sbeamnum}
\end{equation}
averaged over the channel and the integration, where $\Delta\tau_{pq}$ is the per-element residual delay given in  Eq.~\ref{eq:dtau}. This includes near-field defocus, bandwidth smearing and fringe drift, and is expressed in units of the response of one element. The total suppression of RFI in a beam is the product of two factors,
\begin{equation}
\mathcal{S}_{\rm tot} = \mathcal{S}_{\rm PB}(\theta)\;\mathcal{S}_{\rm beam}(\theta,R).
\label{eq:stot}
\end{equation}

\begin{figure}
\centering
\includegraphics[width=\columnwidth]{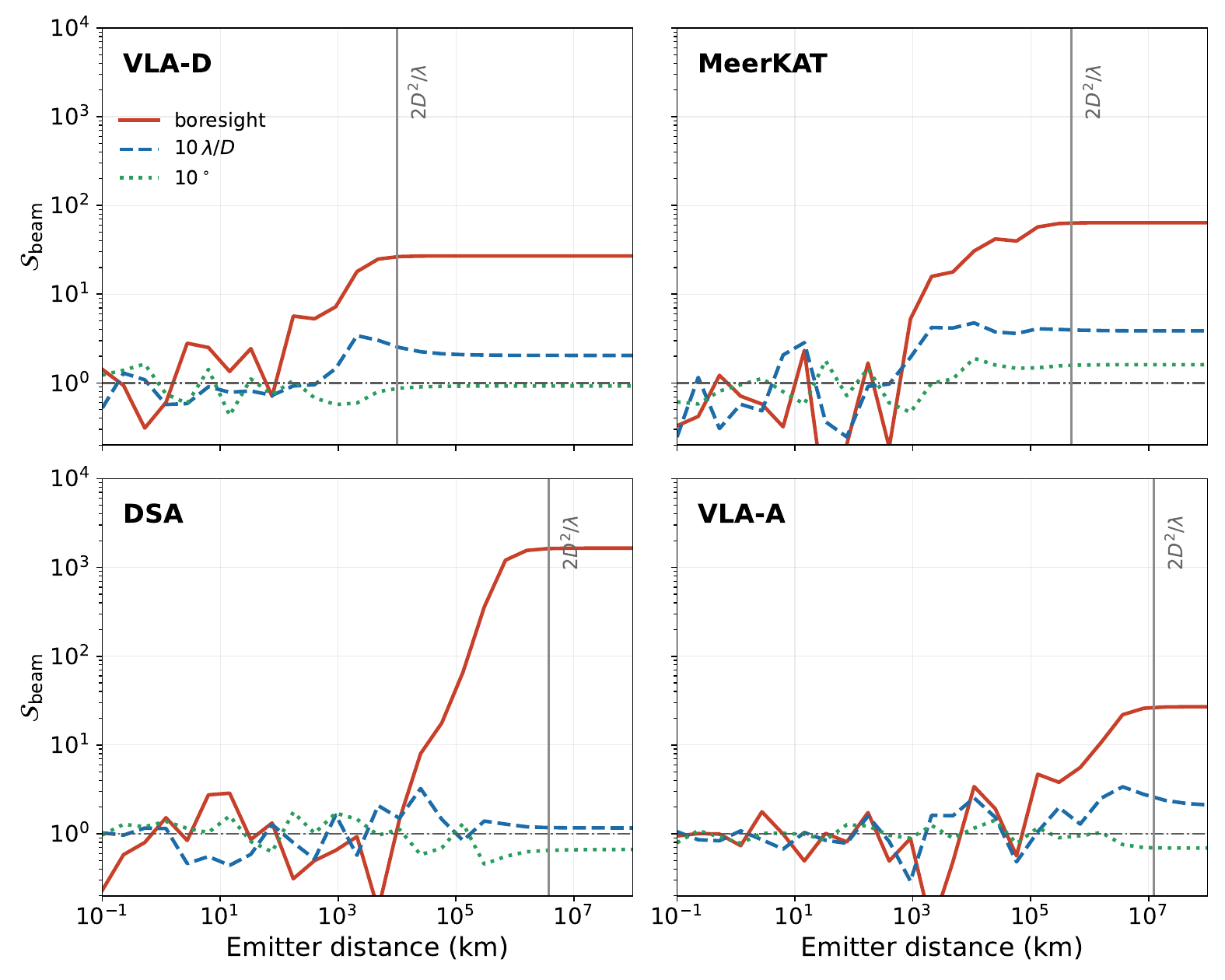}
\caption{Tied-array beam response to a stationary RFI emitter as a function of distance, at three offsets from boresight, for a 1\,MHz channel at 1.4\,GHz. The dot-dashed line marks the incoherent floor $\mathcal{S}_{\rm beam}=1$ and the vertical line indicates the array Fraunhofer distance. At boresight the response saturates at $\mathcal{S}_{\rm beam}=N$. The off-boresight curves are averaged over a small spread of offsets so that they show the mean sidelobe level rather than an accidental null. Curves are computed for a short integration so that they show the geometry alone; a sidereally tracking beam additionally walks off a fixed emitter in $(\lambda/D)/\omega_\oplus$, which is $0.14$\,s for DSA.}
\label{fig:beam}
\end{figure}

Figure~\ref{fig:beam} gives $\mathcal{S}_{\rm beam}$ as a function of emitter range, at boresight, at ten synthesized beamwidths off boresight, and at $10^\circ$. At boresight the response rises from unity in the near field to $N$ in the far field, saturating at the array Fraunhofer distance $2D^2/\lambda$; the transition is at $R\simeq\pi D^2/8\lambda\sqrt N$, which is $380$\,km for VLA-D, $1.4\times10^4$\,km for MeerKAT, $1.9\times10^4$\,km for DSA and $4.7\times10^5$\,km for VLA-A. Ten beamwidths off boresight the coherent response has already collapsed to the sidelobe level, and at $10^\circ$ it is indistinguishable from unity at every range. At terrestrial RFI emitter distances --- below
$\sim\!100$\,km --- every array is at $\mathcal{S}_{\rm beam}\simeq1$ regardless of where the emitter is in the field.

The consequence is that the array factor supplies no suppression of terrestrial RFI. Against a naive
expectation of $1/N$, the measured $\mathcal{S}_{\rm beam}$ at $\theta=45^\circ$ is
within $1.3$\,dB of unity for every array. The reason is the first term of
Eq.~\ref{eq:beam}: with $\sum_p|w_p|^2=1$ the incoherent contribution is $P_{\rm R}$ whatever the phases, so
a coherent beam cannot suppress an emitter below what a single element sees. This motivates beamforming approaches that reject the autocorrelation terms, e.g., with visibility-based beamforming.

For detection, a tied-array beam is the worst place in the instrument to look for RFI, and it remains so for any emitter closer than the Fraunhofer distance, which for a 20\,km array is $3.9\times10^6$\,km, or ten times the lunar distance. The one regime in which a beam detects RFI efficiently is a far-field emitter in the synthesized main lobe, and it is rarely of practical use: the main lobe subtends $\lambda/D$, or $6\times10^{-4}$ degrees for DSA, so an emitter lands in it with probability $\sim\!f/N$.

\section{RFI excision in voltage-data products}
\label{sec:autos}

I have shown that that RFI detection is, for most emitters and most array configurations, optimal in the incoherent combination of per-antenna data. Moving RFI emitters favor the autocorrelation sum for every array considered, by between one and seven orders of magnitude, and faint emitters favor it everywhere. This section addresses how that detection can be realized in practice.

Two properties of voltage-data products are noteworthy. The first is the sensitivity result captured by Eq.~\ref{eq:key}: the autocorrelation sum includes RFI emitters with full coherence and gains $\sqrt N$ from incoherent averaging, with $\mathcal{O}(N)$ computational cost rather than $\mathcal{O}(N^2)$. The second is independent of sensitivity and is arguably more important in practice. Autocorrelations and their sum require $\mathcal{O}(N)$ operations to compute, and they can be formed and tested at whatever time and frequency resolution the emitter requires, rather than at whatever resolution the correlator provides. RFI is often confined in time, in frequency, or in both. For example, a radar transmitting for $1\,\mu$s every millisecond, or a time-division-multiplexed satellite downlink occupying a few percent of the instantaneous spectrum, leaves most of the observation clean but corrupts an entire integration if it is not resolved. For an emitter of duty cycle $\delta$ a matched statistic recovers a factor $\delta^{-1/2}$ over one formed at the integration timescale, which is a factor of 32 for $\delta=10^{-3}$. That gain is available only upstream of the accumulation, and only in a data product cheap enough to form at the required resolution.

A third consideration applies to the highest-severity RFI. Once an emitter drives the signal chain into a non-linear regime, the visibilities acquire a multiplicative error that is degenerate with the antenna gains, and no post-correlation flagging can help. Total power per antenna is the only data in which such an event is directly visible, and monitoring it is the only defense short of avoidance.

I first survey the operations that are performed on voltage-data products (Section~\ref{sec:voltimpl}), then the hardware considerations that determine which of them are feasible in real time (Section~\ref{sec:hardware}), and finally describe a  real-time implementation for the 110-antenna Deep Synoptic Array (DSA-110; Section~\ref{sec:dsa110}).

\subsection{Existing voltage-domain implementations}
\label{sec:voltimpl}

Table~\ref{tab:voltmethods} lists voltage-domain RFI excision techniques in current use, and where they are deployed. They segregate naturally by the data product they operate on rather than by the statistic they compute, and the division is closely aligned with the hardware on which they run. In the table, ``Nyquist'' denotes raw sampled voltages ahead of channelization, and ``ch'' denotes post-filterbank voltages or the power moments formed from them.

\begin{table*}
\centering
\caption{Operations performed on voltage-data products for RFI excision.}
\label{tab:voltmethods}
\begin{tabular}{ccc}
\toprule
Operation (data product) & Targets & Representative implementations \\
\midrule
Threshold blanking  & impulsive broadband &
uGMRT wideband correlator, SPOTLIGHT \\
(Nyquist) & & \citep{Buch2023,spotlight2026} \\[3pt]
Total-power range check & saturation, compression, faults &
DSA, SKA-Mid \\
(Nyquist/ch) & & \citep{Hellbourg2026,gunaratne2019} \\[3pt]
Spectral kurtosis (ch) & narrowband, duty-cycled emitters &
EOVSA, CHIME, DSA-110 \\
& &  \citep{nita2016,taylor2019} \\[3pt]
Median, morphological filtering (ch) & general at high resolution &
Nan\c{c}ay Decameter Array  \\
Time--frequency blanking (Nyquist/ch) & radar, multiplexed downlinks & \citep{lecacheux2013} \\[3pt]
Cross-element outlier tests (autos) & element-specific faults &
DSA-110 \\
Array-summed power (autos sum) & array-common transients & \\[3pt]
Multi-level hierarchical detection & matched to each stage &
UniBoard demonstrator  \\
(Nyquist/ch) & & \citep{dumezviou2016} \\[3pt]
Reference-antenna cancellation  & RFI with reference &
demonstrations  \\
(Nyquist) & & \citep{Kesteven2005,Hellbourg2016} \\
\bottomrule
\end{tabular}
\end{table*}

The operations on Nyquist-sampled voltages are implemented almost exclusively on field-programmable gate arrays (FPGAs), because that is typically where the samples exist before transport and channelization, and because the arithmetic involved is well suited to fixed-point pipelines. The operations on channelized products have migrated to GPUs as correlators themselves have, and are now generally software running on data already resident for the correlator. Total-power range checking appears in many recent large-array designs, reflecting the recognition that compression is not recoverable downstream. The operations in the lower half of the table --- outlier tests across elements, array-summed statistics, and cancellation using a reference --- remain sparsely deployed relative to their apparent value, the last having largely not progressed beyond demonstration.

Operations on raw voltages are the only ones with access to the full time resolution of the instrument, and are consequently the only means to excise the shortest-timescale impulsive broadband RFI. Power-line corona and similar discharges deposit their energy in a few samples implying a flat spread across the band. Once channelized, that energy is distributed over every channel of an integration and is unreachable by any frequency-selective operation. Deployed systems detect such events with a robust location and scale estimator --- typically a median and a median absolute deviation over a sliding block --- and excise samples beyond a fixed multiple of the scale.

Two design choices are common to these implementations, and are worth separating from the detection statistics. The first is what replaces an excised sample. Substituting digital noise of the correct variance preserves the noise statistics, whereas substituting a constant or a threshold value does not. This applies where excision acts on a continuing voltage stream that is subsequently accumulated. Where the flagged samples are instead dropped from a correlation or a sum, zeroing them is the correct treatment, since they are removed from the statistic rather than substituted within it. The replacement noise must be uncorrelated between elements, since a common generator would introduce a spurious correlated signal into every visibility; implementations therefore assign distinct generator states per element \citep{Buch2023}. The second is whether detection is performed on voltage or on power. Voltage-domain detection at Nyquist resolution is typically safe for impulsive astrophysical signals given instrument sensitivities. The same emitter detected in power at millisecond resolution on a large antenna may, on the other hand, be risky. 

Operations on channelized products are correspondingly the means to excise narrowband, potentially duty-cycled RFI. The most widely used statistic is the spectral kurtosis \citep{nita2010}, a normalized ratio of the second and fourth moments of the channel power that tests for departures from the exponential power distribution of Gaussian noise. It requires no bandpass model and is scale-invariant, and is therefore insensitive to arbitrarily shaped instrumental response --- a property that is also its principal limitation, since a steady change in power level is invisible to it. Its expectation is unity, and the sign of the excursion indicates the emitter duty cycle: a continuous signal filling the accumulation depresses it below unity, while an intermittent one lifts it above. The remaining channelized operations in Table~\ref{tab:voltmethods} are outlier tests against a reference, and differ only in what that reference is: neighboring channels of the same element, the same channel on other elements, or the recent history of the array as a whole.

\subsection{Hardware considerations}
\label{sec:hardware}

FPGAs are well matched to RFI excision at the raw voltage stage. Every sample must be processed in order at the sample rate, with little opportunity to buffer and revisit, and an FPGA implements this as a fixed pipeline with deterministic latency. Every sample can receive identical treatment: a blanker that occasionally stalls would impose a time-dependent flag fraction, which is a multiplicative systematic. Integer arithmetic also suits robust statistics: the expensive operation in a median absolute deviation is the median, and for $b$-bit integer samples this is computed by histogram with $2^b$ counters in a single pass, rather than by sorting \citep{Buch2023}. The penalties are that the algorithm is fixed at synthesis, the precision is limited, and operations requiring substantial memory are usually not possible.

GPUs are correspondingly matched to the channelized stage, and increasingly the whole procedure is implemented on GPUs. In systems that transport raw or coarsely channelized voltages to a GPU cluster, all of the operations in Table~\ref{tab:voltmethods} are performed in software on the same data. The data are already resident in device memory for the correlator or beamformer, so the marginal cost of forming power moments alongside is a reduction over data already loaded. That reduction is memory-bandwidth bound rather than arithmetic bound. The thresholds can be derived at run time. This is important for e.g., the spectral kurtosis estimator, where the null distribution is markedly right-skewed at small accumulation depth. A threshold set from a Gaussian quantile misstates the false-alarm rate by more than an order of magnitude, whereas a null distribution computed numerically at startup costs milliseconds.

The design decision that makes a complete chain affordable is to form the moments once. A single pass over the channelized voltages accumulating
\begin{equation}
S_1 = \sum_t |E|^2, \qquad S_2 = \sum_t |E|^4
\label{eq:moments}
\end{equation}
at several accumulation depths simultaneously renders every operation in Table~\ref{tab:voltmethods} a reduction over two resident tensors rather than a second pass over the voltages. Spectral kurtosis follows from $S_1$ and $S_2$ at each depth; the bandpass, cross-element and array-summed tests are all reductions over $S_1$ along different axes. Noise replacement is likewise cheap in this setting, requiring only a per-element generator state to keep the substituted samples mutually uncorrelated.

For current and planned arrays the implication is that the array-summed autocorrelation power is also the easiest to compute, being a single reduction over $S_1$ with no per-baseline work at all. Two conditions determine whether its $\sqrt N$ advantage is realized. It must be formed at the finest accumulation depth available rather than at the correlator integration time, since the duty-cycle gain $\delta^{-1/2}$ is otherwise forfeited. Its thresholds must be calibrated against the realized null rather than assumed, both because the spectral kurtosis null is skewed and because the deviate used in median-based outlier tests is constructed from a location and scale estimated from the same finite sample, so that a nominal $k\sigma$ threshold does not correspond to a $k\sigma$ false-alarm rate. A further consideration, not addressed here, is that the autocorrelation route carries an additive bias, whereas the cross-correlation route does not.

\subsection{DSA-110 implementation}
\label{sec:dsa110}

\subsubsection{Real-time flagger}
\label{sec:dsa110impl}

\begin{figure}
\centering
\includegraphics[width=\columnwidth]{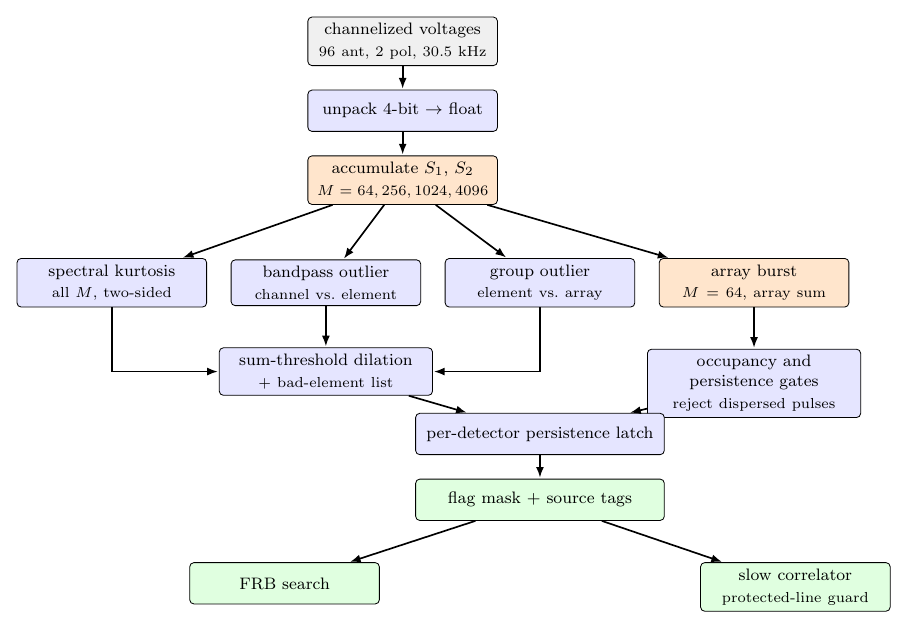}
\caption{Arrangement of the DSA-110 real-time flagger. A single pass over the channelized voltages accumulates the power moments of Eq.~\ref{eq:moments} at four accumulation depths, and all five detectors are reductions over those tensors. The array-burst detector operates at the finest depth and is gated on frequency occupancy, which distinguishes a dispersion-free terrestrial burst from a dispersed astrophysical pulse.}
\label{fig:dsa110}
\end{figure}

The DSA-110 searches for fast radio bursts (FRBs) in real time. RFI excision is required to act on millisecond timescales and to do so without removing the transients being searched for. All operations are performed on channelized voltages  on GPUs, and all detectors are reductions over a single pair of moment tensors. Figure~\ref{fig:dsa110} shows the RFI excision pipeline.

Ninety-six antennas are processed in two polarizations over 187.5\,MHz of bandwidth, channelized to 30.52\,kHz and distributed across nodes in sixteen groups of 384 channels. Voltages are transported as 4-bit complex samples and unpacked on the GPUs (Nvidia RTX2080Ti devices) . The pipeline cadence is a block of 4096 channelized samples, or 134.2\,ms; the channelized sample interval is $32.768\,\mu$s. A single pass over each block accumulates $S_1$ and $S_2$ given in  Eq.~\ref{eq:moments} at four depths simultaneously, $M = 64$, 256, 1024 and 4096 samples, corresponding to 2.097, 8.39, 33.55 and 134.2\,ms. Every detector downstream is a reduction over these tensors.

Five detectors contribute to a combined flag mask, with a per-cell tag recording which was triggered. \emph{Spectral kurtosis} is evaluated two-sided at all four depths against thresholds obtained from a numerically computed null distribution, with the low- and high-side excursions recorded separately so that the emitter duty cycle is preserved in the output. \emph{Bandpass outlier} compares each channel against a robust median bandpass of the same element and polarization; \emph{group outlier} compares each element against the other elements at the same polarization. Both use thresholds calibrated to a specified false-alarm rate against the thermal-noise null hypothesis rather than a fixed multiple of the scatter. \emph{Array burst} operates on the array-summed band power at the finest depth, giving 64 decisions per block, against an exponentially weighted reference over the preceding $\sim\!30$\,s; it is described below. A \emph{sum-threshold dilation} grows isolated detections into contiguous segments. A static list of known bad antennas and a per-detector persistence lock, which holds a cell flagged once the detectors have triggered on it repeatedly, complete the chain. The lock is necessary because an emitter sitting at threshold otherwise blinks at the block rate and leaves partially excised interference behind.

The same flags are consumed by a slow correlator producing spectral-line data, and a fast correlator plus imager pipeline. A configurable guard band excludes a list of protected rest frequencies from flagging on that path. It is required because a detector whose statistic is that a channel is unlike its neighbors cannot distinguish a narrow astrophysical line (usually the 21\,cm HI line) from a narrowband transmitter.

\subsubsection{Short-timescale interference and the array-burst detector}
\label{sec:arrayburst}

The RFI detectors that compare an antenna against a reference share a blind spot for interference that is common to the array, since in that case the reference changes with the signal. For example, a broadband burst lifts every channel of the bandpass reference, and an array-common burst lifts every element of the group reference. Spectral kurtosis is likewise insensitive to a burst that fills whole accumulations, which is Gaussian within them. The array-summed power at the finest available depth is the statistic that catches such bursts, by $\sqrt{N_{\rm a}N_{\rm ch}}$ relative to the single cell the per-element detectors test.

Firing on array-summed power alone would excise the transients the instrument exists to detect, since a bright burst produces the same signature. Two guards separate the two populations. The first applies to RFI filling the sub-band: a sample is flagged only if, in addition to the band-summed power exceeding threshold, at least $90\%$ of the coarse frequency bins are individually occupied. A dispersion-free terrestrial burst fills the sub-band within one 2.097\,ms sample, whereas a dispersed pulse sweeps through it and occupies only a fraction at any instant. The second applies to RFI confined to a few bins, for which occupancy cannot discriminate --- at a dispersion measure of several hundred pc\,cm$^{-3}$ a pulse sweeps only a few bins per sample, comparably to a band-limited impulse. There the discriminant is persistence: a pulse visits each bin once as it sweeps, whereas interference revisits the same bins, so a bin is excised only if it has also fired in several of the preceding blocks. That history is accumulated causally, from strictly earlier blocks, and only from blocks that were themselves band-limited, so that a bright source recurring within the history window cannot arm its own excision.

Flagged cells are zeroed in the voltages ahead of correlation rather than replaced, so that every baseline touching a flagged element contributes nothing on the affected channels. The per-block and array-burst masks are combined into a single multiplicative factor to avoid a second pass over the voltage block.

\subsubsection{Example results}

\begin{figure}
    \centering
    {(\textit{a})}
    \includegraphics[width=0.8\textwidth]{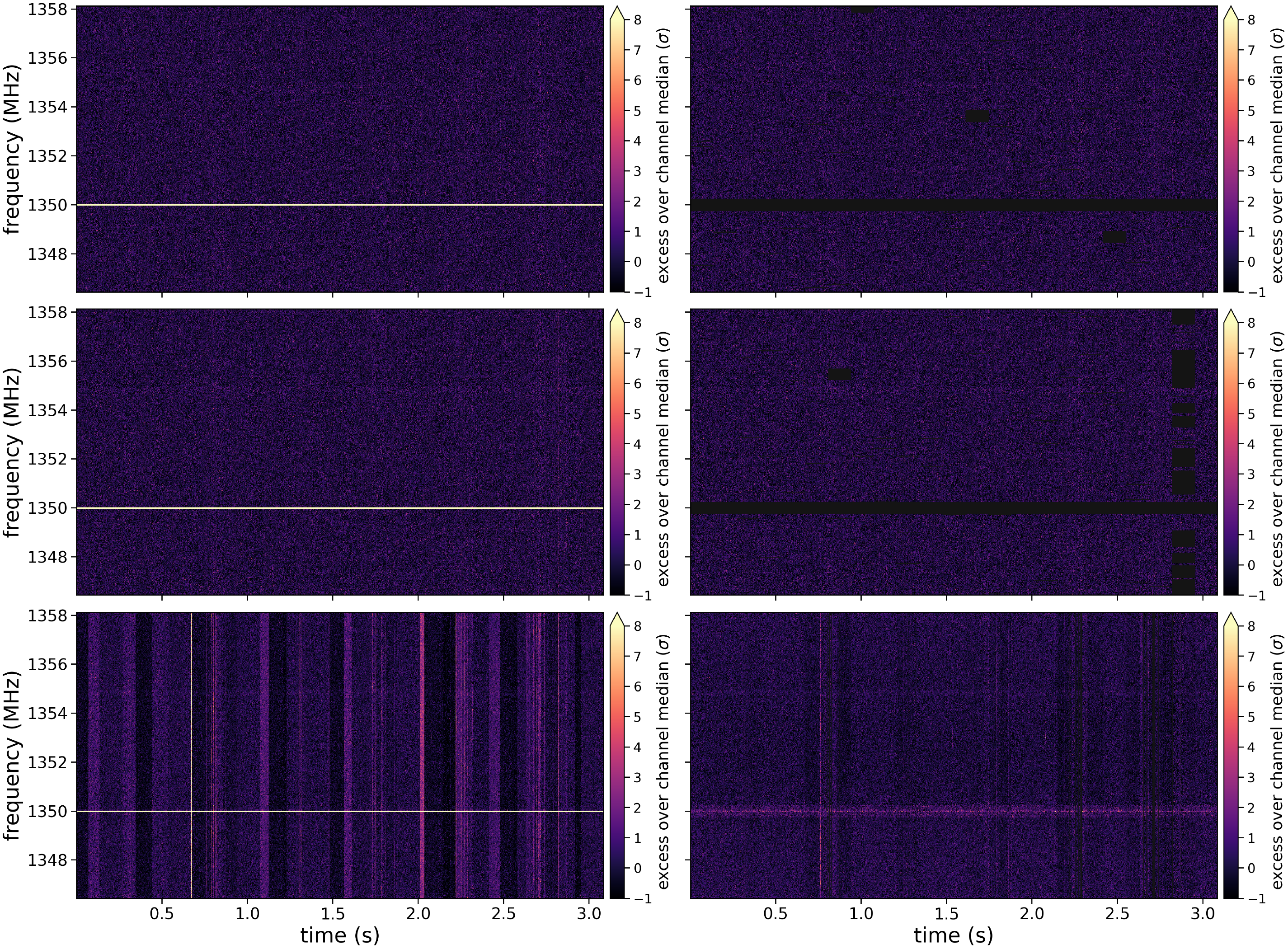} \\
    {(\textit{b})}
    \includegraphics[width=0.8\textwidth]{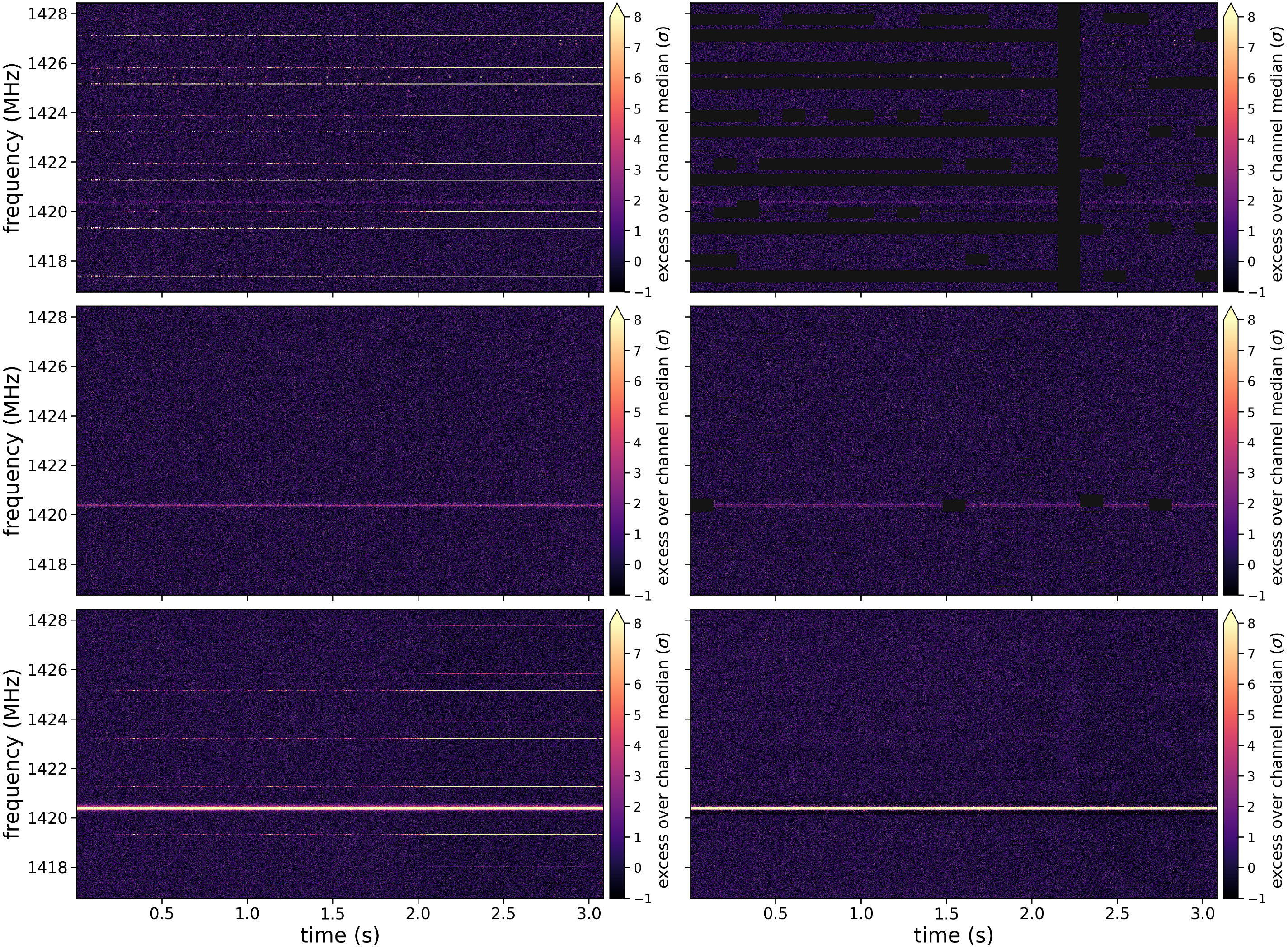}
    \caption{Two examples of RFI excision in DSA-110 voltage data. For each of panel-sets (a) and (b), the top two rows show dynamic spectra of data from two DSA-110 antennas spaced by 400\,m, and the third row shows dynamic spectra of data summed over 48 antennas of the EW arm of the DSA-110 core. The left columns show data prior to real-time RFI excision, and the right columns show the cleaned data. Panel-set (a) has examples of broadband impulsive RFI that are primarily evident in the array sum, although some bursts are evident (and flagged) in the middle row. Panel-set (b) has examples of time-variable narrow-band RFI. Note that the 21\,cm HI line is intentionally not excised.} \label{fig:voltage}
\end{figure}

\begin{figure}
    \centering
    \includegraphics[width=0.8\textwidth]{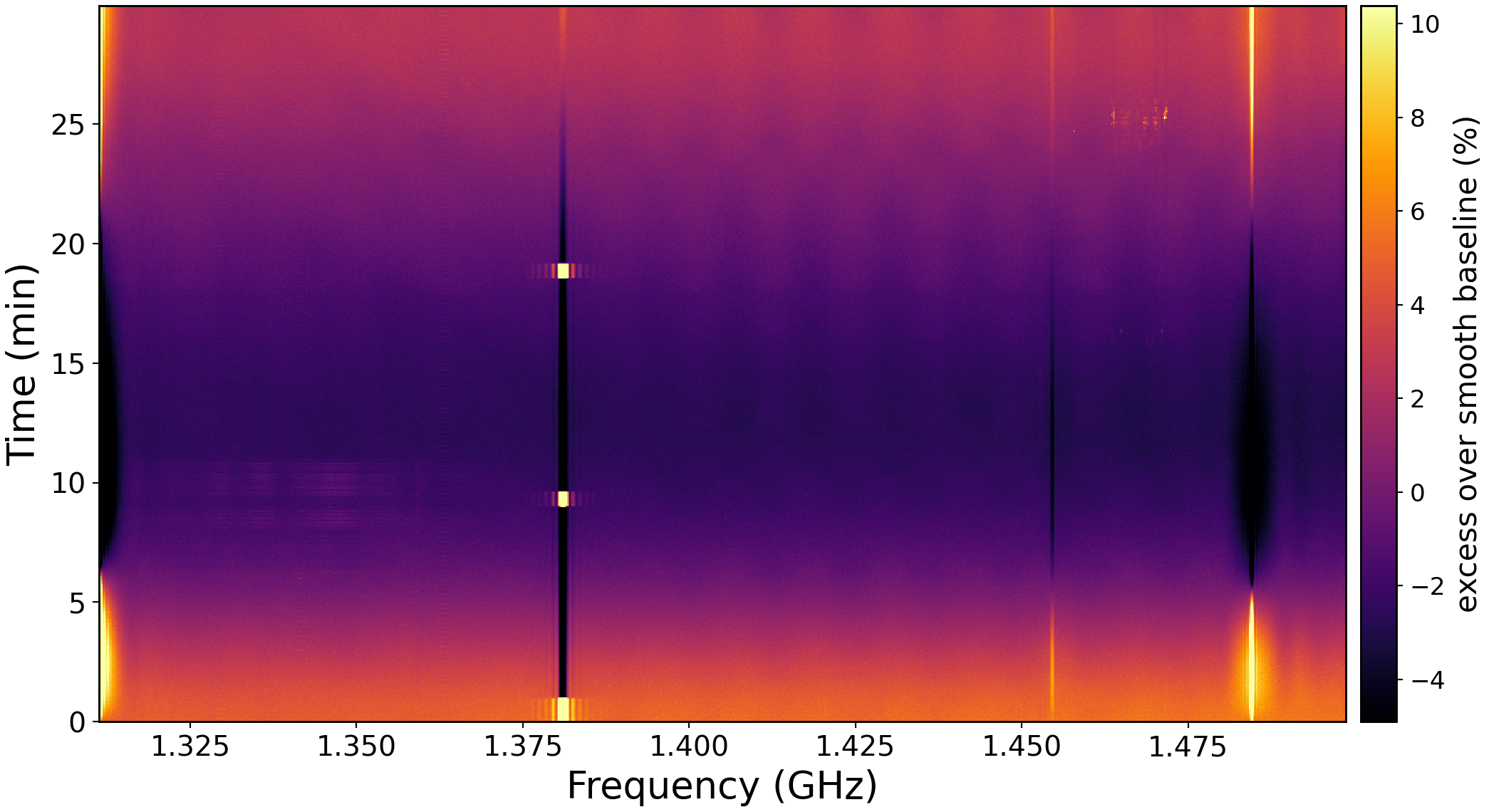} \\
    \includegraphics[width=0.8\textwidth]{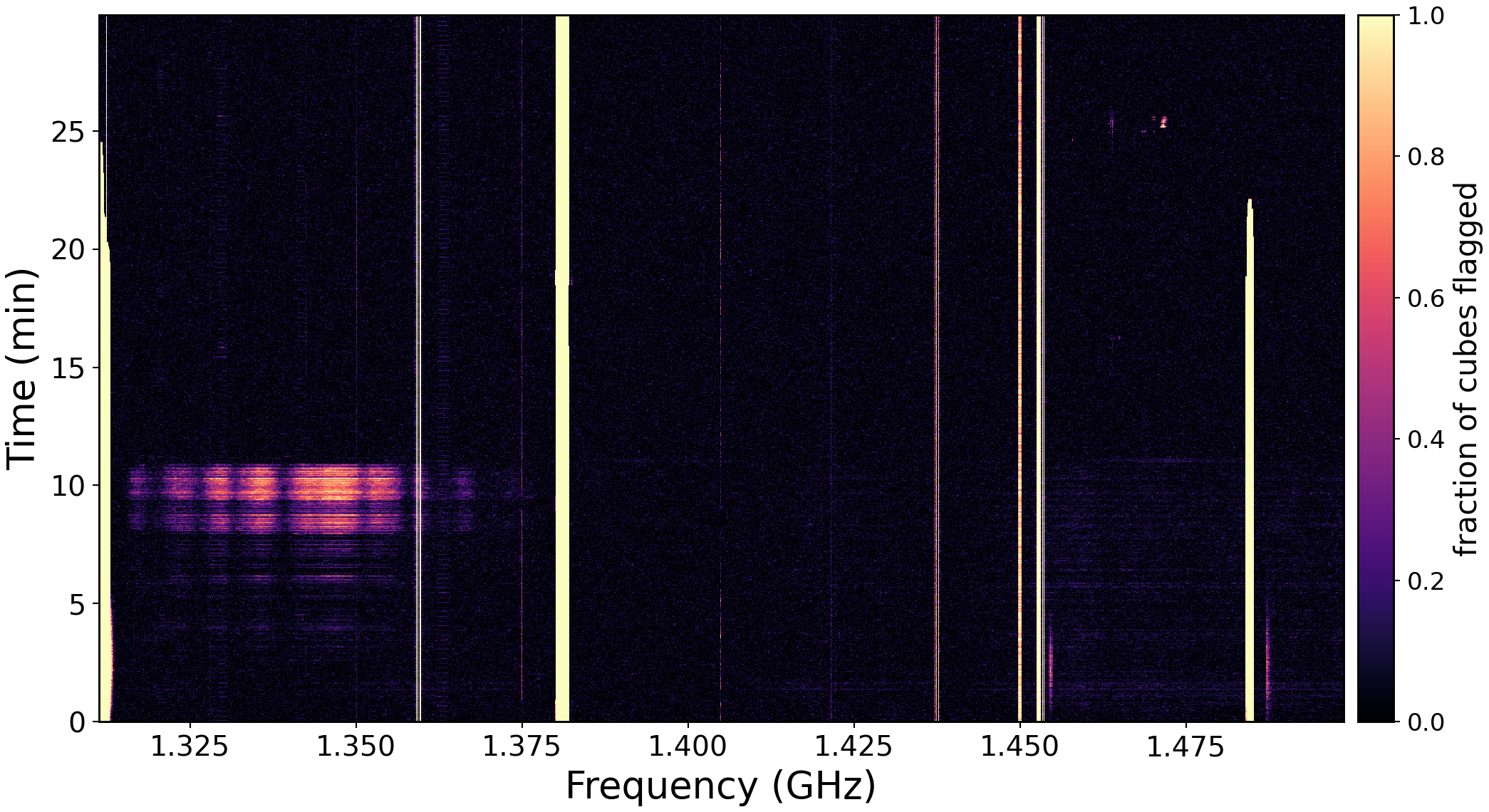} 
    \caption{\textit{Top}: time-integrated dynamic spectrum of data from a single DSA-110 antenna, showing examples of RFI. \textit{Bottom}: waterfall plot for the same section of time showing the percentage of real-time flagged data per display pixel for this antenna. } \label{fig:waterfall}
\end{figure}

Examples of the performance of the DSA-110 RFI excision scheme are shown in Figures~\ref{fig:voltage} and \ref{fig:waterfall}. The scheme is implemented as part of the \texttt{dsa110-rt} package\footnote{\url{https://github.com/dsa110/dsa110-rt}}, within a processing stage that computed millisecond-timescale dedispersed visibilities per sub-band. The scheme runs at $\sim10\times$ real-time in operation. This RFI excision is sufficient to minimize the FRB trigger rate to a manageable level above the detector threshold in the fast image-plane search (event rate of $\sim3$\,hr$^{-1}$). 

Figure~\ref{fig:voltage} shows examples of the RFI excision scheme run on triggered voltage dumps with the excision  turned off. The top panel set shows examples of narrow-band and impulsive broadband RFI, and the bottom panel set shows examples of time-variable narrow-band RFI. Both cases are reasonably well excised. Analysis of the different RFI sources evident in DSA-110 data is ongoing. 

Figure~\ref{fig:waterfall} shows typical diagnostic plots over 30\,min of operations for a single polarization of a single antenna. The waterfall-plot spectra are integrated to 2.147\,s, and RFI-excision percentages over this timescale are shown in the bottom flag-waterfall plot. RFI of various classes are excised, although some signals (e.g., sidebands of bright impulsive $\sim1380$\,MHz RFI) are not fully flagged. 

\section{Discussion}
\label{sec:discussion}

The goal of RFI excision is not the same in every application, and the appropriate operating point follows from the data product and the science case. Fast-transient searches require control of both Type I and II error rates, since a false positive enters a candidate list that must be vetted and a false negative is a missed burst. Pulsar timing and continuum imaging instead require data free of systematics at the thermal-noise limit, which tolerates aggressive flagging but not biased flagging. A threshold that preferentially removes positive noise excursions biases flux densities whether or not it removes any interference. Spectral-line observations add the requirement that bandwidth be preserved. The detection step is common to all three, and it is the step this paper addresses.

Section~\ref{sec:strategies} reduces that step to a single comparison: the advantage of a cross-correlation statistic over the autocorrelation sum is the RFI coherence times the square root of the number of antennas, with a crossover at $\langle\gamma d\rangle = (N-1)^{-1/2}$, or $0.20$ for a 27-element array, $0.13$ for 64 and $0.025$ for 1650. Emitters in motion fall one to seven orders of magnitude below the autocorrelation sum on every configuration considered, because their coherence is destroyed by rapid fringe rotation rather than by array geometry, and no choice of algorithm recovers it without shorter averaging times being implemented. Bright stationary terrestrial emitters favor the cross-correlations on compact and on filled arrays, but not on extended sparse ones. For RFI below the single-antenna noise the phase-unknown cross-correlation statistic does not outperform the autocorrelation sum in any configuration examined.

In practice this argues for a hierarchical arrangement rather than a single choice of data product. Individual antenna streams should be flagged and conditioned first, and the autocorrelation sum inspected afterwards. The ordering matters because the sum is not robust to the conditions per-element tests are designed to find: a saturated or faulty element, or interference confined to a subset of the array, enters the sum as an excess indistinguishable from a genuine array-common signal, and dilutes the $\sqrt N$ advantage that motivates forming it. Conditioning first leaves the sum to simply find low-level array-common signals the per-antenna excision cannot remove.

The same ordering helps with gain stability, which is the principal systematic limiting the per-antenna route and which I have otherwise neglected. The autocorrelation carries an additive bias whereas the cross-correlation does not, and the sum is bound by the fractional stability of the common-mode gain rather than the radiometric limit. Antenna to antenna fluctuations average down in the sum, making it more robust than any individual autocorrelation; common-mode fluctuations do not, and set the bound. I am not aware of a published measurement of common-mode gain stability on any array at the relevant timescales.

Excision after correlation or beamforming remains necessary in several instances. Compact arrays observing stationary emitters preserve coherence almost perfectly, and there the cross-correlations are both the moresensitive detector and the place the RFI does most damage. Recovering the full $\sqrt{N_{\rm b}}$ advantage requires the emitter phase, however, and without prior information that means a search over trial positions. The statistical penalty for such a search is mild, growing only as the logarithm of the number of trials, but the cost of performing it on every time--frequency cell in real time is not. Untargeted searches are therefore impractical for most systems, favoring incoherent array statistics of the kind introduced by \citet{ssins}.

It is increasingly imperative that excision be incorporated into real-time processing. The immediate driver is the move towards delivering highly reduced, science-ready data products \citep{dsa2000,srdp}: an excision step that cannot be performed before the data are averaged and archived cannot be performed at all. A second is that both the interference environment, and some astronomical signals of interest, are becoming more complex in their temporal behavior, and the matched statistic for a signal confined in time exists only at resolutions available prior to accumulation --- a factor $\delta^{-1/2}$ for an emitter of duty cycle $\delta$, or 32 for $\delta=10^{-3}$, forfeited entirely by a detector operating at the correlator integration time. The corresponding benefit is the recovery of clean time. RFI that is resolved can be excised surgically rather than by discarding whole integrations.

However, any excision applied in real time modifies the array selection function, and the modification must be monitored and recorded. Observing time, $uv$ coverage and effective bandwidth are all altered by flagging, and a downstream analysis assuming any of them uniform will be wrong in a way that is difficult to diagnose afterwards. Flag provenance --- which detector was triggered, on which cells, at what threshold --- should be carried alongside the flags. GPUs are well suited to this. Strategies can be implemented and revised in software, thresholds calibrated against the realized null at run time rather than assumed, and the data are already resident in device memory for the correlator or beamformer. Inspection of pre-accumulation data is consequently computationally minor: forming the power moments once renders every subsequent test a reduction over resident tensors, and the array-summed statistic identified in Section~\ref{sec:strategies} as the most sensitive is also the cheapest to compute.

The optimal pathway depends on the array, on the local RFI environment and on the science cases, and careful analysis for specific instruments is warranted. The treatment herein is intended to demonstrate how such an analysis may proceed rather than to prescribe a result. The inputs are the array configuration, the emitter geometry and the choices of channel width and integration time, and the output is a ranking of data products by detection margin. The broader point is one of framing. Most treatments of RFI mitigation ask how best to detect interference in a given data stream, whether visibilities, beamformed intensities or voltages. I have argued that it is equally useful to ask which data stream a given emitter is best detected in, and that the answer is frequently not the one in which the excision is ultimately applied.

\section{Conclusions}
\label{sec:conc}

I have analyzed the detectability of different forms of RFI in different data products of radio-telescope arrays. Special attention is given to the trade-off between detection in data streams from individual antennas, treated individually or power-summed, and in post-correlation and beamformed data. I present a framework for performing such analyses, and do so for a selection of representative RFI emitters and radio arrays. Although suppression of RFI in certain data products may be seen as beneficial, I adopt the approach that low-level RFI that nonetheless affects performance near the noise floor or detection threshold of an instrument may be more easily identified and excised if the right data product is inspected. In a rapidly evolving and increasingly crowded RFI environment, such analyses are critical in determining how best to allocate signal-processing resources in radio arrays to mitigating RFI to produce scientifically useful data. I conclude the following.

\begin{enumerate}

\item RFI signals of various types are suppressed in visibility and tied-array beam data. RFI emitters within approximately an array diameter are suppressed due to near-field wavefront curvature in visibilities, and significant suppression is observed within an array Fraunhofer distance for tied-array intensities. Prevalent moving RFI emitters, in particular aircraft and LEO/MEO satellites, are significantly suppressed in fringe-tracked visibilities for most large sparse arrays. 

\item In many cases, the autocorrelation sum is more sensitive to RFI than other commonly used RFI detection schemes: individual baseline flagging, combinations over baselines (without actual coherent RFI searches), and tied-array beams. The autocorrelation sum, prefaced by excision in individual-antenna data streams to remove antenna-specific RFI, has the advantages of ease of computation, and access to time- and frequency-resolutions that can be matched to RFI sources. I present heuristics (e.g., Eq.~\ref{eq:key}) to identify the regimes where the autocorrelation sum is most useful. It is of course critical that the autocorrelation sum be optimally formed from the individual antennas, and that the effects of gain fluctuations do not significant affect RFI detection. 

\item Real-time signal processing systems in modern radio telescopes, typically implemented on FPGA and GPU hardware, are well suited to implementing a variety of RFI excision tools. I urge the designers of current and upcoming radio-array digital backend systems to approach the design of RFI excision from the whole-of-system perspective taken here, identifying which RFI detection schemes are optimal for the specific array configurations, and RFI emitters of interest. I present a real-world demonstration of a real-time pre-correlation RFI detector that operates on GPUs in the DSA-110 backend. 

\end{enumerate}

\section*{Acknowledgments}

I thank Greg Hellbourg and Liam Connor for comments on this manuscript. This material is based in part upon work supported by the National Science Foundation under Grants No. 2537086 and 1836018. This research is enabled in part by development funded by Schmidt Sciences for the Deep Synoptic Array project (DSA). The DSA is part of the Eric and Wendy Schmidt Observatory System in partnership with the California Institute of Technology. The code used to conduct the analyses and produce the figures in this manuscript was prepared with significant assistance from Claude (Fable and Opus models). The code was manually checked, and tests performed, to ensure that the intent of the author was represented. The manuscript was drafted by the author with the assistance of Claude (Fable models) in polishing the text.

\bibliographystyle{plainnat}

\bibliography{oja}

\end{document}